\documentclass[showpacs,aps,pre,twocolumn,floatfix,superscriptaddress]{revtex4-1}

\usepackage{graphicx}
\usepackage{epstopdf}
\usepackage{dcolumn}
\usepackage{array}
\usepackage{gensymb}
\usepackage{bm}
\usepackage{setspace}
\usepackage{color}
\usepackage{graphicx}
\usepackage{epstopdf}
\usepackage{dcolumn}
\usepackage{bm}
\usepackage{gensymb}
\usepackage{mathcomp}
\usepackage{textcomp}
\usepackage{setspace}
\usepackage{amsmath}
\usepackage{amsthm}
\usepackage{amssymb}
\usepackage[capitalize]{cleveref}

\newcommand{\bI}{\mathbb{I}}

\newcommand{\ep}{\varepsilon}

\newcommand{\sqX}{{\widetilde{X}}}
\newcommand{\sqY}{{\widetilde{Y}}}
\newcommand{\sqW}{{\widetilde{W}}}
\newcommand{\sqS}{{\widetilde{\Sigma}}}

\newcommand{\sqtau}{\widetilde{\tau}}
\newcommand{\vu}{\vec{u}}
\newtheorem{lemma}{Lemma}

\begin{document}
\title{Why Capillary Flows in Slender Triangular Grooves \\
  Are So Stable Against Disturbances}
\author{Nicholas C. White and Sandra M. Troian}
\homepage[Corresponding author; URL:~]{stroian@caltech.edu; www.troian.caltech.edu}
\affiliation{California Institute of Technology, T. J. Watson Sr. Laboratories of Applied Physics, 1200 E. California Blvd., MC 128-95, Pasadena, CA 91125, USA}
\date{\today}

\begin{abstract}
Ongoing development of fuel storage and delivery systems for space probes, interplanetary vehicles, satellites and orbital platforms continues to drive interest in propellant management systems that utilize surface tension to retain, channel and control flow in microgravity environments. Although it has been known for decades that capillary flows offer an ideal method of fuel management, there has been little research devoted to the general stability properties of such flows. In this work, we demonstrate theoretically why capillary flows which channel wetting liquids in slender open triangular channels tend to be very stable against disturbances. By utilizing the gradient flow form of the governing fluid interface equation, we first prove that stationary interfaces in the presence of steady flow are asymptotically nonlinearly and exponentially stable in the Lyapunov sense. We then demonstrate that fluid interfaces exhibiting self-similar Washburn dynamics are transiently and asymptotically linearly stable to small perturbations. This second finding relies on a generalized non-modal stability analysis due to the non-normality of the governing disturbance operator. Taken together, these findings reveal the robust nature of transient and steady capillary flows in open grooved channels and likely explains the prevalent use of capillary flow management systems in many emerging technologies ranging from cubesats to point-of-care microfluidic diagnostic systems.
\end{abstract}
\date{\today}


\maketitle

\section{Introduction}
\label{sec:introduction}
\begin{figure}
\centering
\includegraphics[scale=1.0]{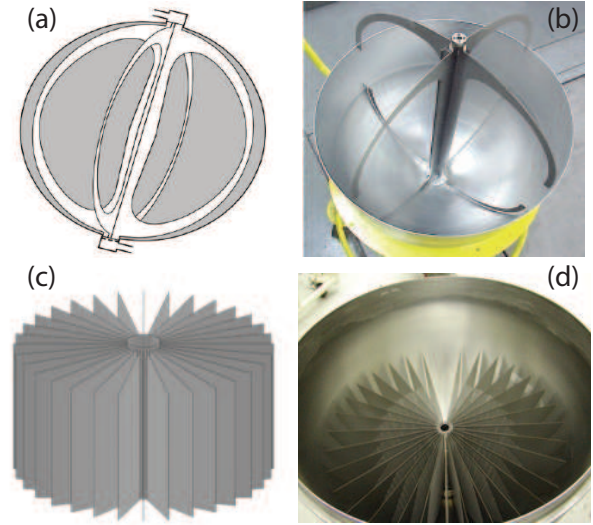}
\caption{(Color online) Two sample propellant management devices (PMDs) for insertion into satellite fuel tanks for routing liquid within open grooved channels. (a) Sketch and (b) fabricated structure showing vane type PMD. (c) Sketch and (d) fabricated structure showing sponge type PMD. Images (a), (c) and (d) from www.pmdtechnology.com - copyright \copyright 2011 PMD Technology. Image (b) reprinted from Ref. [\onlinecite{J91}] with permission from the American Institute of Aeronautics and Astronautics.)}
\label{fig:PMDsamples}
\end{figure}

It has been known for centuries that wetting liquids will rapidly and spontaneously creep along surfaces containing grooves, interior corners, crevices or roughened areas, a process known as wicking. Since the late 1960s, researchers have been incorporating this passive and reliable method of flow control in the design of novel propellant management devices able to store, channel and meter fuel resourcefully in microgravity environments \cite{RJ85,J91,LJ15,H16,H17}. Such systems have significantly extended mission lifetimes of spacecraft and satellites, enabling future interplanetary explorations as well. Modern propellant management systems consist of combinations of sponges, traps, troughs, vanes, and wicks to channel propellant flow by capillary action, systems which have been investigated extensively \cite{J91,J97,WDec01,WC02,CC10,DC17}. Shown in Fig. \ref{fig:PMDsamples} are two common structures designed in such a way that the liquid film thickness is much smaller than the streamwise flow distance, a limiting ratio which as described in Section \ref{sec:flowbackground} leads to considerable simplification of the governing equations of motion. However, despite that by the 1920's the mechanism describing internal capillary flow, whereby a liquid column spontaneously fills the interior of a slender capillary tube, was understood and the appropriate equations developed, the mechanism driving spontaneous capillary flow along an open grooved channel required a half-century more to be deduced. Ongoing efforts to miniaturize fluid management systems for many different applications continue to drive interest in the fundamentals of free surface capillary flow along structured substrates.

Edward Washburn appears to have provided the first theoretical analysis of the distance in time traversed by a Newtonian liquid flowing within a horizonal enclosed cylindrical tube under the action of capillary forces \cite{W21}. The Washburn relation, as it is now known, is given by the relation $z = (\gamma \, \cos \theta\, R \, t/ 2 \mu)^{1/2}$, where $z$ is the distance traversed by the advancing (for wetting fluids) or receding (for non-wetting fluids) meniscus, $t$ is time, $R$ is the cylinder radius, and $\mu$, $\gamma$ and $\theta$, respectively, are the liquid shear viscosity, surface tension and contact angle. Although this largely unidirectional flow is driven by the capillary pressure drop across the curved meniscus separating gas from liquid, the Washburn relation can be regarded from a mathematical point of view as an example of diffusive driven interface growth since the ``diffusion'' coefficient preceding the time variable is described by units of length squared per unit time. By treating porous bodies as an assemblage of small cylindrical capillaries, Washburn also went on to predict that the volume of liquid penetrating a porous medium should scale in time as $(\gamma t / \mu)^{1/2}$, where geometric factors and wettability constants were incorporated into an overall proportionality constant. Washburn thereby revealed the essential physical mechanism governing the spontaneous creep of liquids into enclosed spaces, a process that can easily dominate and oppose the force of gravity for sufficiently small enclosures. The Washburn scaling has since been used successfully to quantify capillary transport in many different systems ranging from blood flow in microvascular hemodynamics \cite{J69,SS89} to oil extraction from porous rocks \cite{WM76} to water uptake in dried tree specimens \cite{JK10}, to name a few. Through modern day advances in microfabrication techniques, capillary action is now being incorporated into the design of many microfluidic devices as well, some involving spontaneous flow through chemically  treated porous substrates, others relying on a combination of capillary action, positive displacement pumping and electrophoresis. The number of such applications is multiplying rapidly with emphasis on disposable inexpensive platforms beneficial to global public health \cite{YW06}, drug discovery screening \cite{DM06} and specialized fluid based logic circuitry \cite{TQ02,PG07}.

While Washburn originally focused on the fluid dynamics underlying internal capillary flow, there soon came the realization that capillary forces were also somehow responsible for the spontaneous exterior flow of liquid along \emph{open} surfaces manifesting  grooves, edges, corners and roughened patches. This type of free surface flow is typically characterized by a solid boundary whose surface pattern is fixed and a gas/liquid interface whose shape undergoes spatial and temporal variation as the flow evolves toward steady state conditions. Free surface wicking is now also being incorporated into many small-scale fluidic devices such as heat pipes for cooling of microelectronics \cite{CL84,MW92,PW93,QS17}, surface-enhanced Raman spectroscopic devices \cite{PM07} and small spacecraft fluid management systems \cite{CW09}.
\begin{figure}
\centering
\includegraphics[scale=0.7]{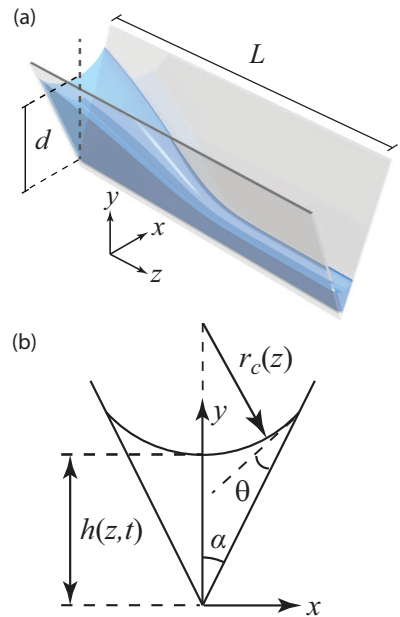}
\caption{(Color online) (a) Schematic of a wetting liquid film ($0 \leq \theta < \pi/2$) flowing within a slender open triangular groove with constant cross-section. The inlet midline film thickness at the origin is denoted by $d=h(x=0,z=0,t)$ and the channel length by $L$ where $d/L \ll 1$. (b) Cross-sectional view of the flow geometry depicted in (a) where $h(z,t)$ denotes the local midline film thickness, $\alpha$ is the groove interior half-angle, $r_c$ is the radius of curvature of the liquid interface and $\theta$ is the contact angle of the liquid wetting the channel sidewalls, which is assumed constant.}
\label{fig:vgroove_diagram}
\end{figure}

\section{Background}
\label{sec:background}
In one of the earliest studies of free surface wicking of a Newtonian liquid, Concus and Finn \cite{CF69} in 1969 showed that a liquid drop will penetrate the interior of an open triangular channel with opening half-angle $\alpha$ so long as $\theta + \alpha < \pi /2$. Shortly thereafter, Ayyaswamy et al. \cite{AE74} derived and numerically solved for the stationary (i.e. steady state) streamwise velocity field for this geometry. Closed-form analytical solutions for the free surface transient capillary flow in idealized  containers with grooves or corners were demonstrated two decades later by several groups working independently. Dong and Chatzis \cite{DC95}, Weislogel and Lichter \cite{W96,WL98} and Romero and Yost \cite{RY96a} derived and solved for the interface equation $h(z,t)$ describing the thickness of a liquid film flowing within an open slender triangular groove satisfying the condition $d \ll L$, as sketched in Fig. \ref{fig:vgroove_diagram}. The mechanism responsible for spontaneous capillary flow within an open groove turned out to be more complex than its counterpart for small enclosures. While for interior flow the liquid column is pulled forward by the net force of surface tension acting along the triple line (i.e. the line defining the junction between the gas, liquid and solid media), the free boundary in an open capillary driven system experiences a normal force across the entire interface whose shape typically varies in space and time. By implementing two key assumptions, namely that the  streamwise variation in liquid height varies smoothly on a length scale $L$ much larger than $d$, and that the flow is slow enough such that capillary forces dominate inertial and viscous forces, these researchers were able to show that the cross-section of the liquid surface must describe a circular arc of radius $R(z)$. Hence the pressure gradient driving the flow stems exclusively from the streamwise variation in $R(z)$. These two assumptions therefore led to a nonlinear diffusion equation describing the relaxation of the cross-sectional liquid area proportional to $h^2(z,t)$. While Romero and Yost \cite{RY96a} and Dong and Chatzis \cite{DC95} assumed unidirectional flow as well, Weislogel and Lichter \cite{W96,WL98} were able to derive similar results without  relying on this simplification by instead conducting a formal perturbation expansion  based on the slender parameter approximation $(\varepsilon)^2 = (d/L)^2 \ll 1$. These separate efforts led to the conclusion that there exist transient solutions according to which the penetration distance $z$ scales in time as $(D \, t)^{1/2}$ where $D=\gamma \, d \, K(\theta, \alpha)/\mu$, $d$ is a characteristic liquid thickness (such as the groove depth or midline liquid thickness) and $K(\theta, \alpha)$ is a geometric factor. Chen, Weislogel, and Nardin \cite{CN06} later generalized this result to channels with constant cross-section ranging from the nearly-rectangular to highly rounded grooves. In all cases, it was confirmed that the penetration distance retains the $t^{1/2}$ scaling. Numerous experimental studies \cite{RY96b,RO98,W96,CW09,DS14,BP15} and references therein have since confirmed their predictions.

A few studies have gone beyond solution of the flow base states to explore their stability spectrum. Weislogel \cite{W01} examined the linear stability of an infinite, initially quiescent column of liquid supported within a slender triangular groove of constant cross-section using conventional modal analysis. Other researchers have since examined the free surface collapse and gas ingestion in inertial-capillary flows for rectangular \cite{HD09}, triangular \cite{WH13} or asymmetrical \cite{TH15} grooves, the stability of liquid menisci attached to a solid edge \cite{L90,YH07}, and the dryout instability for thermocapillary (i.e. non-isothermal) driven flow in triangular grooves \cite{YH06}. Aside from these specialized studies, however, the majority of theoretical studies have been devoted exclusively to solutions governing unperturbed base states i.e. solutions in the absence of disturbances.

In this work, we address for the first time the stability of stationary and transient interfaces associated with capillary flows in slender, open triangular grooves of constant cross-section subject to the Concus-Finn inequality \cite{CF69} ensuring liquid penetration ($\theta + \alpha < \pi /2$) and the restriction that the liquid always wet the channel sidewalls. In Section \ref{sec:flowbackground}, we outline the derivation of the governing interface equation \cite{RY96a,DC95,W96,WL98}. In Section \ref{sec:steadystate}, we examine the nonlinear stability of stationary states subject either to Dirichlet, Neumann or constant volume conditions. By invoking the gradient flow form of the interface disturbance equation and an associated Lyapunov energy functional governing the decay in time of arbitrary disturbances, we demonstrate that stationary interface shapes are exponentially stable and therefore extremely robust to perturbations. In Section \ref{sec:selfsimilar}, we examine the generalized non-modal linear stability of self-similar states obeying Washburn scaling \cite{TD93,FI96} using both analytic arguments and direct numerical computation. Our results indicate that self-similar solutions characterized by the scaling $z \sim t^{1/2}$ are both transiently and asymptotically stable to infinitesimal perturbations.

\section{Model for free surface capillary flow in slender open triangular grooves}
\label{sec:flowbackground}
In this section, we outline the theoretical model describing the capillary motion of a non-volatile, isothermal Newtonian liquid film flowing within an open slender triangular groove in contact with an ambient passive gas \cite{DC95,RY96a,W96,WL98}, as depicted in Fig. \ref{fig:vgroove_diagram}(a). The inlet midline film thickness at the origin is denoted by $d=h(x=0,z=0,t)$ and the channel length by $L$ where the slender limit is assumed, namely $d/L \ll 1$. Shown in Fig. \ref{fig:vgroove_diagram}(b) is a cross-sectional view of the flow geometry where $h(z,t)$ denotes the local midline film thickness, $\alpha$ is the groove interior half-angle, $r_c$ is the radius of curvature of the liquid interface and $\theta$ is the contact angle of the liquid wetting the channel sidewalls. The liquid film is assumed to maintain a constant contact angle  independent of location or flow speed.

\subsection{Slender limit form of the hydrodynamic equations}
In order to conserve mass and momentum, an incompressible Newtonian liquid of constant density must satisfy the continuity and Navier-Stokes equation given by the set of coupled equations
\begin{align}
\label{eqn:Dcontinuity}
\nabla \cdot \vu &= 0, \\
\label{eqn:DNS}
\rho \left[\frac{\partial \vu}{\partial t} + \left( \vu \cdot \nabla \right)
\vu \right] &= -\nabla p + \mu \nabla^2 \vu + \rho \vec{g},
\end{align}
where the velocity field in Cartesian coordinates is represented by $\vec{u} = (u,v,w)$, the fluid pressure by $p(x,y,z)$, the gravitational acceleration by $\vec{g}$ and the constant fluid density by $\rho$. Assuming flow in a slender channel such that  $\varepsilon = d/L \ll 1$ and a dominant balance between the pressure gradient and the viscous force per unit volume leads to the characteristic scalings and non-dimensional variables listed in Table \ref{table:nondim} along with the corresponding Bond $\textsf{Bo}$, capillary $\textsf{Ca}$ and Reynolds $\textsf{Re}$ numbers in the slender limit.
\begin{table}
\centering
\begin{tabular}{l l l}
\hline
\hline
&& \\
~~~~~Quantity & Scaling & Rescaled \\
&& variable\\
\hline
&& \\
\, Slender parameter & & \, $\varepsilon = d/L \ll 1$ \\
&& \\
\, Coordinates & $x_c = d$ & \,$X = x/x_c$ \\
& $y_c = d$  & \,$Y = y/y_c$\\
& $z_c = L$ & \,$Z = z/z_c$ \\
&& \\
\, Velocity & $u_c = \varepsilon^2 \gamma \textsf{Ca}/\mu$ & \,$U = u/u_c$ \\
& $v_c = \varepsilon^2 \gamma \textsf{Ca}/\mu$ & \,$V = v/v_c$ \\
& $w_c = \varepsilon \gamma \textsf{Ca}/\mu$ & \,$W \!= w/w_c$ \\
&& \\
\, Pressure & $p_c = \gamma \textsf{Ca} / (\varepsilon L)$ & \,$P = p/p_c$ \\
&& \\
\, Time & $t_c = \mu L / (\varepsilon \gamma \textsf{Ca})$ & \,$T = t/t_c$ \\
& & \,$\tau = \ln(T)$ \\
&& \\
\, Interface midline & $y_c = d$ & \,$H = h(z,t)/y_c$ \\
\,\,\,\,\, thickness &&\\
\, Interface shape & $y_c = d$ & \,$\Sigma = \sigma(x,z,t)/y_c$ \\
\, Interface radius & $y_c = d$ & \,$R = r_c/y_c$ \\
\,\,\,\,\, of curvature &&\\
&& \\
\, Stationary state & & \,$H_S (Z)$ \\
\,\,\,\,\,  midline thickness &&\\
&& \\
\, Self-similar variable & &\, $\eta = Z/\sqrt{T}$ \\
\, Self-similar state& &\, $S(\eta) $ \\
\,\,\,\,\, midline thickness && \\
&& \\
\, Bond number & & \,$\textsf{Bo}=\rho g d^2/\gamma$ \\
\, Capillary number & & \,$\textsf{Ca}=\mu w_c/\varepsilon \gamma=\Phi$ \,\\
\, Reynolds number & & \,$\textsf{Re}= \rho w_c d/\mu$ \,\\
&& \,\,\,\,\,\,\,\,\,$=(\varepsilon \rho \gamma d/\mu^2) \textsf{Ca}$ \,\\
&& \\
\hline
\hline
\end{tabular}
\caption{Characteristic scalings (lower case) and non-dimensional variables (uppercase) used to describe dimensionless system shown in Fig. \ref{fig:vgroove_diagram}.}
\label{table:nondim}
\end{table}
To order $\varepsilon^2$, the rescaled forms of Eqs. (\ref{eqn:Dcontinuity}) and (\ref{eqn:DNS}) are then given by
\begin{subequations}
\begin{align}
0 &=\frac{\partial U}{\partial X} + \frac{\partial V}{\partial Y} + \frac{\partial W}{\partial Z}, \\
\ep^3 \textsf{Re} \frac{D U}{D T} &=\frac{\textsf{Bo}}{\textsf{Ca}} G_x-\frac{\partial P}{\partial X} + \ep^2 \Delta U,\\
\ep^3 \textsf{Re} \frac{DV}{D T} &=\frac{\textsf{Bo}}{\textsf{Ca}} G_y
- \frac{\partial P}{\partial Y} + \ep^2 \Delta V, \\
\ep \textsf{Re} \frac{D W}{D T} &=\frac{1}{\ep} \frac{\textsf{Bo}}{\textsf{Ca}} G_z-\frac{\partial P}{\partial Z} + \Delta W,
\end{align}
\end{subequations}
where the substantial derivative $D/DT$ and the Laplacian derivative $\Delta$,  respectively, are given by
\begin{subequations}
\begin{align}
\frac{D}{D T} &= \frac{\partial}{\partial T} + U\frac{\partial}{\partial X} + V \frac{\partial }{\partial Y} + W \frac{\partial}{\partial Z},\\
\Delta &=\frac{\partial}{\partial X^2} + \frac{\partial}{\partial Y^2} + \ep^2 \frac{\partial}{\partial Z^2} ,
\end{align}
\end{subequations}
and $\vec{G} = \vec{g}/g$. In the limits where $\varepsilon^2 \ll 1$, $\varepsilon \, \textsf{Re} \ll 1$ and $\textsf{Bo}/\textsf{Ca} \ll \varepsilon$, the governing equations reduce to the form:
\begin{subequations}
\begin{align}
\label{eqn:continuity_final}
\frac{\partial U}{\partial X} + \frac{\partial V}{\partial Y} + \frac{\partial W}{\partial Z} &= 0, \\
\label{eqn:PXYconstant}
\frac{\partial P}{\partial X} = \frac{\partial P}{\partial Y} &= 0,\\
\frac{\partial P}{\partial Z}
&= \frac{\partial^2 W}{\partial X^2} + \frac{\partial^2 W}{\partial Y^2} . \end{align}
\end{subequations}
When subject to the slender limit, the flow is therefore inertia-free and the fluid pressure is constant throughout the $(x,y)$ plane. The pressure gradient driving the flow, which will stem solely from capillary forces, can therefore only vary along the  streamwise axis and can only be counterbalanced the viscous force set in play by the  no-slip boundary condition applied along the groove sidewalls, namely $U=V=W=0$ at all liquid/solid interfaces.

\subsection{Boundary conditions at the liquid interface}
The two (dimensional) boundary conditions specifying the jump in normal and shear stresses across the gas/liquid interface $\sigma(x,z,t)$ are given by
\begin{subequations}
\begin{align}
\left[\hat{n} \cdot (\sqtau - p \bI) \cdot \hat{n} + \gamma(\nabla_s \cdot \hat{n})
\right ]_{y=\sigma(x,z,t)} &=  0,\\
\left[\hat{t}_{i=1,2} \cdot \sqtau \cdot \hat{n}
\right]_{y=\sigma(x,z,t)}
&= 0,
\end{align}
\end{subequations}
where $\bI$ is the 3$\times$3 identity matrix,
$\sqtau = \mu [\nabla \vu + (\nabla \vu )^T]$ denotes the shear stress tensor, $\nabla_s = (\nabla - \hat{n} (\hat{n} \cdot \nabla))$ denotes the surface gradient operator and the triad ($\hat{n},\hat{t}_1,\hat{t}_2$) denotes the three unit vectors representing directions normal and tangent to the moving interface with the convention that $\hat{n}$ points away from the interface. In rescaled units, these unit vectors are given by
\begin{subequations}
\begin{align}
\hat{N}
&= \frac{1}{[1 + (\partial_X \Sigma)^2 + \ep^2 (\partial_Z \Sigma)^2 ]^{1/2}}
\begin{pmatrix}
-\partial_X \Sigma \\
1 \\
-\ep \partial_Z \Sigma
\end{pmatrix},
\\
\hat{T}_1
&= \frac{1}{[1 + (\partial_X \Sigma)^2]^{1/2}}
\begin{pmatrix}
1 \\
\partial_X \Sigma \\
0
\end{pmatrix},
\\
\hat{T}_2
&= \frac{1}{[1 + \ep^2 (\partial_Z \Sigma)^2]^{1/2}}
\begin{pmatrix}
0 \\
\ep \partial_Z \Sigma \\
1
\end{pmatrix},
\end{align}
\end{subequations}
where $\Sigma(X,Z,T) = \sigma(x,z,t)/d$ denotes the non-dimensional interface function and subscripts denote differentiation with regard to the rescaled coordinates. Specifying a system for which the fluid pressure derives solely from variations in the local interface curvature of the flowing liquid, the interfacial surface tension $\gamma$ is everywhere constant since the liquid is isothermal and contains no surfactant-like additives, and that the liquid remains in contact with a passive quiescent gas of negligible viscosity and density with gauge pressure set to zero, the jump in normal stress is then strictly due to capillary forces and the liquid interface is a surface of vanishing shear stress. To order $O(\varepsilon^2)$ then, these dimensionless boundary conditions reduce to the form
\begin{subequations}
\begin{align}
\label{eqn:Peqn}
0
&= - P - \textsf{Ca}^{-1} \left(\frac{\partial^2_X \Sigma}{[1 + (\partial_X \Sigma)^2]^{3/2}} \right) + O(\varepsilon^2)\\
\label{eqn:NormalBC}
&= - P - \textsf{Ca}^{-1}\textsf{K}(Z,T) + O(\varepsilon^2),\\
0
&= \partial_Y W - (\partial_X \Sigma) \partial_X W + O(\varepsilon^2), \\
\notag
0
&=\left[ 1 - (\partial_X \Sigma)^2 \right](\partial_Y U + \partial_X V) \\
\notag
& \phantom{=}+ 2 (\partial_X \Sigma) (-\partial_X U + \partial_Y V)
\\
\label{eqn:ShearBC}
& \phantom{=} - (\partial_Z \Sigma)\left[\partial_X W + (\partial_X \Sigma) \partial_Y W \right] + O(\varepsilon^2),
\end{align}
\end{subequations}
where $\textsf{K(Z)}$ represents the local curvature of the interface function $\Sigma$ (defined to be positive for a wetting liquid). It is clear from Eq. (\ref{eqn:NormalBC}) that the curvature function $\textsf{K}$ can in general only depend on $(Z,T)$ since according to Eq. (\ref{eqn:PXYconstant}), the pressure $P$ is independent of $(X,Y)$. This then requires that the cross-sectional shape of the liquid interface be described by a curve with constant curvature. This restriction limits the shape either to a flat interface or one described by a segment of a circle. Since the liquid must also satisfy a  prescribed contact angle set by the particulars of the liquid/solid interaction, a flat profile is disallowed and $\Sigma(Z,T)$ must therefore trace out a circular arc of constant curvature. The non-dimensional radius of curvature of the gas/liquid interface is then given by $R(Z,T)= H(Z,T)\,\sin\alpha/(\cos\theta - \sin\alpha)$, or likewise, the interface curvature is described by $K(Z,T) = 1/R(Z,T) = (\cos\theta - \sin\alpha) (\csc\alpha) H^{-1}(Z,T)$. (This relation differs slightly from that originally derived by Romero and Yost \cite{RY96a} where they adopted a sign convention for $\textsf{K}$  opposite to ours and chose the reference liquid thickness to be the height of the fluid intersecting the groove wall, not the inlet midline film thickness.) According to Eq. (\ref{eqn:Peqn}), the capillary pressure is then given by
\begin{align}
\label{eqn:pressure_height}
P(Z,T) = &- \frac{\textsf{Ca}^{-1}}{\widehat{R}(\alpha,\theta)H(Z,T)},
\end{align}
where
\begin{equation}
\widehat{R}(\alpha,\theta) = \frac{\sin\alpha}{\cos\theta - \sin\alpha} .
\end{equation}
The Concus-Finn condition $\alpha + \theta < \pi/2$ for liquid imbibition yields $\widehat{R}(\alpha,\theta) > 0$, or likewise $P(Z,T) < 0$, consistent with a liquid  interface with positive curvature. The case $\alpha + \theta > \pi/2$ is not relevant to our study since it ultimately leads to dewetting configurations resulting in a cascade  instability resembling a linear array of primary, secondary, and tertiary droplets   \cite{YH07}).

\subsection{Interface midline equation $H(Z,T)$ for capillary flow in slender open triangular grooves}
\begin{figure}
\centering
\includegraphics[scale=0.8]{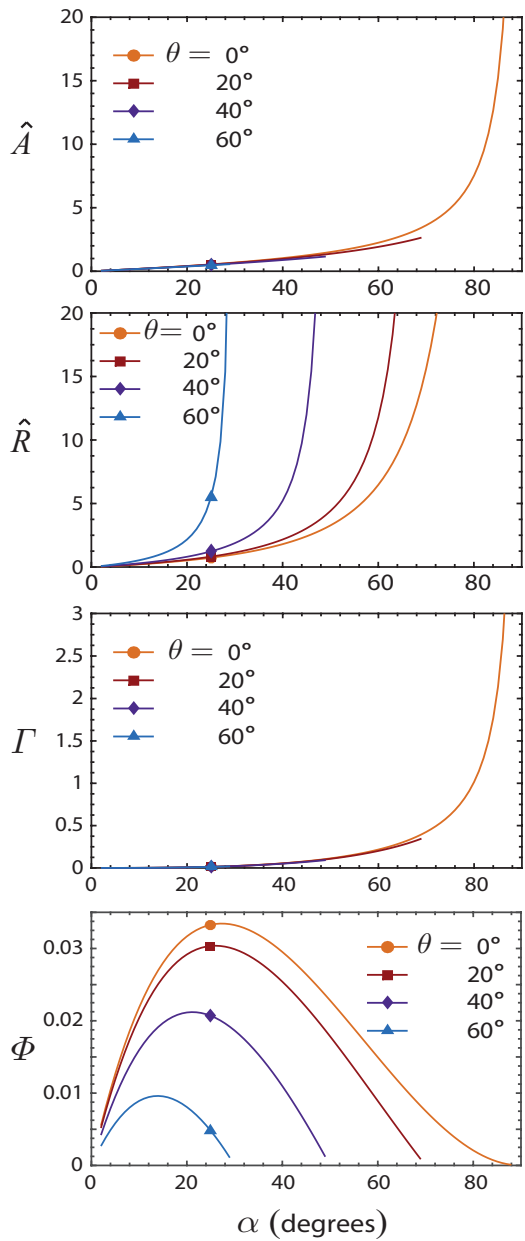}
\caption{(Color online) Geometric functions pertinent to capillary flow of a Newtonian liquid film with constant contact angle $\theta$ in a slender open triangular groove with half opening angle $\alpha$ satisfying the Concus-Finn condition $\alpha + \theta < \pi/2$. Plotted are the functions (a) $\widehat{A}(\theta,\alpha)$, (b) $\widehat{R}(\theta,\alpha)$, (c) $\Gamma(\theta, \alpha)$ and (d) $\Phi(\theta, \alpha)$ described in the text.}
\label{fig:geometric_factors}
\end{figure}

The fact that the interface shape can only be a segment of a circle, and is therefore independent of the local coordinates $(X,Y)$, leads to simplification of the expression for the streamwise volumetric flux $Q(Z,T)$. The relevant variables are then scaled by $H$ according to
\begin{align}
\sqX
&=\frac{X}{H} \quad \textrm{and} \quad \sqY = \frac{Y}{H}, \\
\sqW(\sqX, \sqY)
&= \left(- \frac{\textsf{Ca}^{-1}}{\widehat{R}} \, \frac{\partial H}{\partial Z}\right)^{-1} W, \\
\sqS(\sqX)
&= 1 + \widehat{R} - \sqrt{\widehat{R}^2-\sqX^2}.
\end{align}
This rescaling allows for solution of $\sqW$ independently of the local value of $H$. As a result, the geometric function $\Gamma(\alpha,\theta) = \iint \sqW d\sqX d\sqY$ need only be computed numerically once. The dimensionless streamwise flux which traverses the local cross-sectional area $A$ can then be re-expressed as
\begin{align}
\label{eqn:NondimFlux}
Q(Z,T)
&=  \iint_A W~dXdY \\
&= - \frac{\textsf{Ca}^{-1}}{\widehat{R}(\alpha, \theta)} \Gamma(\alpha,\theta) H^2 \frac{\partial H}{\partial Z},
\end{align}
where the integrated area $A = \widehat{A}(\alpha,\theta) H^2$ and
\begin{equation}
\widehat{A} = \frac{\cos\theta \sin\alpha\cos(\alpha + \theta) - \left(\pi/2 - \alpha - \theta\right) \sin^2\alpha}{(\cos\theta - \sin\alpha)^2}
\end{equation}
is a geometric factor \cite{W96,RY96a}. Since the local streamwise gradient in liquid flux is directly related to the time derivative of the local liquid cross-sectional area (see Appendix 2 of Ref. \onlinecite{LZ84} for derivation) according to $\partial A/\partial T = -\partial Q/\partial Z$, the governing nonlinear diffusion equation for the midline height $H(Z,T)$ is then given by
\begin{equation}
\label{eqn:master_nowc}
\widehat{A}(\alpha,\theta) \frac{\partial H^2}{\partial T}
= \frac{\textsf{Ca}^{-1}}{\widehat{R}(\alpha,\theta)} \Gamma(\alpha,\theta) \frac{\partial}{\partial Z} \left( H^2 \frac{\partial H}{\partial Z} \right).
\end{equation}
Without loss of generality and to recast this equation into parameter-free form, the remaining scaling for the streamwise velocity is chosen to be $w_c = (\varepsilon \gamma / \mu) \Phi$, where $\Phi(\alpha,\theta)= \textsf{Ca}$ (see Table 1) is defined by
\begin{equation}
\Phi(\alpha,\theta) = \frac{\Gamma(\alpha,\theta)}{\widehat{A}(\alpha,\theta)\,\widehat{R}(\alpha,\theta)} .
\end{equation}
The resulting interface equation, whose stability properties we examine in this work, is then given by
\begin{equation}
\label{eqn:H2Master}
\frac{\partial H^2}{\partial T} - \frac{\partial}{\partial Z} \left( H^2
\frac{\partial H}{\partial Z} \right)  = 0,
\end{equation}
subject to the constraint that $H(Z,T)$ is everywhere always positive. In Fig. \ref{fig:geometric_factors} are plotted the scaled functions $\widehat{A}(\alpha,\theta)$, $\widehat{R}(\alpha,\theta)$, $\Gamma(\alpha,\theta)$ and $\Phi(\alpha,\theta)$ for four values of the liquid contact angle $\theta = 0^\circ$, $20^\circ$, $40^\circ$ and $60^\circ$ as a function of increasing groove interior half-angle $\alpha$. While $\widehat{A}(\alpha,\theta)$, $\widehat{R}(\alpha,\theta)$ and $\Gamma(\alpha,\theta)$ are of order $O(10)$ or less, the values of $\Phi(\alpha,\theta)$ are far smaller and tend toward $O(10^{-2})$ or less. Note too that since $\widehat{R}(\alpha,\theta)$ and $\Gamma(\alpha,\theta)$ are non-negative functions for systems obeying the Concus-Finn condition, the direction of the liquid flux specified by Eq. (\ref{eqn:NondimFlux}) is then strictly determined by the sign of the local interface slope $\partial H/ \partial Z$. Interfaces with $\partial H/ \partial Z < 0$ engender a positive local flux $Q$ and vice versa. A vanishing local flux results whenever $\partial H/ \partial Z = 0$.

The form of Eq. (\ref{eqn:H2Master}) falls within a class known as the porous media equation generally given by $\partial C(Z,T)/\partial T = \partial^2 C^m /\partial Z^2$, where $C(Z,T)$ is a non-negative scalar function and $m$ is a constant larger than one \cite{N84,R84,O01,V07}. As discussed in Ref. [\onlinecite{V07}], this nonlinear diffusion equation describes the relaxation of the order parameter $C(Z,T)$ relevant to various phenomena which arise in different branches of science and mathematics and exhibiting properties such as scale invariance and self-similarity. To help track the energy flow associated with the evolution of $C$, Newman \cite{N84} outlined a general method for constructing Lyapunov functionals for systems sustaining traveling wave solutions. He used that method to establish the actual rate of convergence of an initial configuration to solutions exhibiting self-similarity. Ralston \cite{R84} complemented this work by showing that Newman's choice of Lyapunov function allowed proof that initial conditions with finite mass converge to self-similar solutions asymptotically in time. When applied to our system, these results indicate that self-similar states which result from the spreading of an initial finite drop within a slender open triangular groove are asymptotically globally stable. However, despite longstanding interest in the porous media equation and wicking phenomena in general, there has been no prior study of which we are aware of the general stability of stationary or transient solutions corresponding to unconstrained volume states.

\section{Stability of stationary interface solutions $H_S (Z)$}
\label{sec:steadystate}
In this section, we first derive the analytic form corresponding to stationary solutions of the governing interface equation given by Eq. (\ref{eqn:H2Master}) with geometric factors explicit. By invoking analogy to the porous media equation and constructing an appropriate Lyapunov functional, we then examine the nonlinear asymptotic stability of these states against arbitrary disturbances.

\subsection{Stationary states for time-independent Dirichlet, Neumann and volume conditions}
\label{StatSolns}
Stationary states of Eq. (\ref{eqn:H2Master}) correspond to those solutions for which $\partial H^2/ \partial T = 0$, which reduces the governing equation to a second order, ordinary differential equation requiring two boundary conditions. So long as the interface slope does not vanish, such stationary interfaces $H_S(Z)$ are possible because the subsurface flow of liquid establishes a balance between the local capillary and viscous stresses generates a constant flux $Q_o$, which according to Eq. (\ref{eqn:master_nowc}) is given by
\begin{align}
\label{eqn:QS}
Q_o
&= - \frac{\textsf{Ca}^{-1}}{\widehat{R}(\alpha,\theta)} \Gamma(\alpha,\theta) H^2_S \frac{dH_S}{dZ} \\
&= - \widehat{A}(\alpha,\theta) H^2_S \frac{dH_S}{dZ}.
\end{align}
The general form of these stationary solutions is then represented by
\begin{equation}
\label{eqn:H_SGeneral}
H_S = \left[ \textsf{const} - \frac{3 Q_o Z}{\widehat{A}(\alpha,\theta)} \right]^{1/3} > 0\,
\end{equation}
described by a power law decrease of $Z^{1/3}$ for positive flux and power law increase for negative flux. Particular solutions, of course, require specification of two boundary conditions for setting the values of \textsf{const} and $Q_o$. For a triangular channel with fixed groove opening angle and liquid contact angle (i.e. constant value of $\widehat{A}(\alpha,\theta)$) extending between endpoints $Z_1$ and $Z_2$, stationary solutions $H_{S}$ correspond to
\begin{equation}
\label{eqn:steadycuberoot_fluxdir}
H_S = \left[H^3_1 - \frac{3 Q_S}{\widehat{A}(\alpha,\theta)} (Z - Z_1) \right]^{1/3}
\end{equation}
for Dirichlet and Neumann conditions $H_S (Z_1) = H_1$ and $Q_2 = Q_S$, respectively.
Alternatively, Dirichlet conditions imposed at both endpoints, such that $H_S(Z_1) = H_1$ and $H_S(Z_2) = H_2$, yield
\begin{equation}
\label{eqn:steadycuberoot}
H_S(Z) = \left[H_2^3 \frac{Z-Z_1}{Z_2-Z_1} + H_1^3 \frac{Z_2-Z}{Z_2-Z_1} \right]^{1/3} .
\end{equation}
Note from Eq. (\ref{eqn:pressure_height}) and the scaling for the streamwise flow speed $w_c$ that specification of the boundary film thickness is equivalent to specification of the boundary fluid pressure. Solutions $H_S$ to Eq. (\ref{eqn:H_SGeneral}) can also be generated subject to constant flux $Q_S$ at one boundary ($Q_1=Q_S$ or $Q_2=Q_S$) and conservation of volume $V_S$, which sets the constant value $C_S$ in the implicit relation :
\begin{align}
\label{eqn:steadycuberoot_fluxvol}
V_S
& = \widehat{A} \int_{Z_1}^{Z_2} H^2 (Z) dZ \\
& = \frac{3^{5/3} \widehat{A}^{1/3}}{5 Q_S}
\left[ (C_S - Q_S Z_1)^{5/3}- (C_S - Q_S Z_2)^{5/3} \right] .
\end{align}
Likewise, constant volume and fixed liquid height at one endpoint yield similar forms.

\begin{figure}
\centering
\includegraphics[width = 8.0 cm]{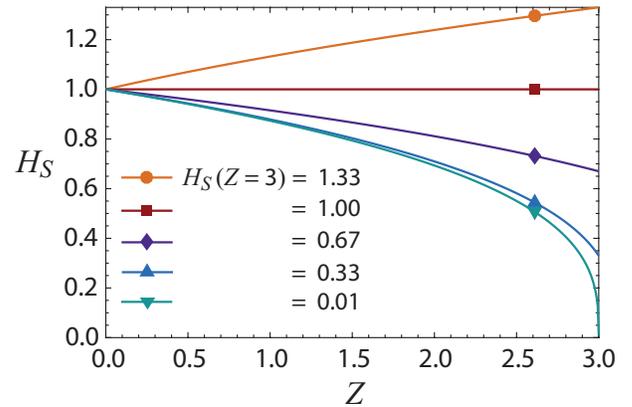}
\caption{(Color online) Representative stationary solutions subject to Dirichlet conditions $H_S(Z_1) = H_1 = 1.0$ and $H_S(Z_2)= H_2$ = 0.01, 0.33, 0.67, 1.00 and 1.33 for the range $0 \leq (Z_1, Z_2) \leq 3.0$.}
\label{fig:steady_solutions}
\end{figure}

Representative solutions $H_S(Z)$ are plotted in Fig. \ref{fig:steady_solutions} for  Dirichlet conditions which pin the inlet height to $H_1 = 1.0$ and pin the outlet height $H_2$ to the five values shown. From the expression for the flux given by Eq. (\ref{eqn:QS}), it is evident that the solution with $H_2 = 1.33$, which exhibits interface slope values $dH/dZ$ which are everywhere positive, describes a stationary solution with net streamwise flux $Q_S < 0$ i.e. net flow directed from right to left. The uniform solution $H_2 (Z) = 1.0$ clearly then represents a case with no flux and no subsurface flow i.e. a quiesence liquid filament. The remaining curves with negative interface slopes throughout the domain correspond to stationary solutions with positive flux i.e. net flow directed from left to right. Other boundary conditions yield similar shapes with characteristic scaling $Z^{1/3}$.

\subsection{Nonlinear exponential stability of stationary solutions}
\label{sec:nonlinstab}
We next examine the time-asymptotic nonlinear stability of stationary solutions by appealing to a Lyapunov analysis. In particular, we construct a Lyapunov energy function (akin to a potential energy function in classical mechanics) to determine whether and how rapidly arbitrary disturbances equilibrate back to the stationary solutions $H_S$. A similar approach to asymptotic stability has previously been used to characterize disturbance decay governed by nonlinear diffusion dynamics \cite{KF77}. Here we show that the capillary flow in a bounded domain with time-independent boundary conditions has a uniquely determined stationary solution which is dynamically stable. This implies that any initial distribution which maintains the condition $H(Z,T)>0$ for all time will ultimately evolve toward the stationary state. Furthermore, it is shown that initial states satisfying Dirichlet boundary conditions are \textit{globally stable} - that is, any initial distribution $H(Z,0)>0$ will converge to the stationary state. For all sets of boundary conditions posed, we show that the convergence to the stationary state is exponential in time so long as it satisfies the positivity condition $H(Z,T)>0$.

To begin, it proves convenient to recast Eq. (\ref{eqn:H2Master}) into gradient flow  form \cite{CT94,GO01}
\begin{equation}
\frac{\partial G(Z,T)}{\partial T} - \frac{\partial}{\partial Z} \left\{ M(G) \frac {\partial}{\partial Z}\left(\frac{\delta \mathfrak{F}}{\delta G}\right) \right \}
= 0 ,
\label{eqn:GradientForm}
\end{equation}
where $M(G)$ is the so-called mobility function and $\delta \mathfrak{F}/\delta G$ denotes the functional derivative of $\mathfrak{F}(H)$ with respect to $G$. The quantity  $\mathfrak{F}(H)$ represents the Lyapunov free energy $\mathfrak{F}$ which for \textit{stable} stationary states must satisfy the relations $\delta \mathfrak{F} = 0$ (i.e. extremal condition) and $\delta^2  \mathfrak{F} > 0$ (positivity condition on Hessian for stable flow). By introducing $G(H) = H^2(Z,T)$ and letting $M = 1/3$, Eq. (\ref{eqn:H2Master}) is re-expressed as
\begin{equation}
\label{eqn:hyperbolic_form}
\frac{\partial G}{\partial T} = \frac{\partial}{\partial Z} \left[ \frac{1}{3} \frac{\partial R}{\partial Z}  \right]
\end{equation}
where the function $R(G)$ is given by
\begin{equation}
\label{eqn:R(Z)}
R(G) = ( G^{3/2} - G^{3/2}_S ) = (H^3 - H^3_S) .
\end{equation}
(The mapping from $H$ to $G$ is one-to-one since $H$ is strictly positive.) Comparison with Eq. (\ref{eqn:GradientForm}) then yields the correspondence $\delta \mathfrak{F}(G)/\delta G = R(G)$ subject to the constraint that $[R \partial R/\partial Z]_{Z_1,Z_2} = 0$. This definition of $R(G)$ ensures that stationary states are identified by the extrema of the free energy $\mathfrak{F}(H)$ where  $\left. \delta \mathfrak{F}(G)/\delta G \right|_{G_S}= R(G_S)=0$. Additionally, Dirichlet boundary conditions correspond to $R|_{Z_1,Z_2}=0$ and Neumann boundary conditions to $(\partial R/\partial Z)|_{Z_1,Z_2} = 0$. Multiplication of both sides of Eq. (\ref{eqn:hyperbolic_form}) by $R(G)$ followed by integration over the finite domain $[Z_1, Z_2]$ yields the relation
\begin{align}
\notag
\int_{Z_1}^{Z_2} R \frac{\partial G}{\partial T}dZ &
= \int_{Z_1}^{Z_2} \frac{R}{3} \frac{\partial}{\partial Z} \left[ \frac{\partial R}{\partial Z} \right]dZ
\\
\notag
&= \left[ \frac{R}{3} \frac{\partial R}{\partial Z} \right]_{Z_1}^{Z_2} - \int_{Z_1}^{Z_2} \frac{1}{3} \left[ \frac{\partial R}{\partial Z} \right]^2 dZ
\\
&= - \int_{Z_1}^{Z_2} \frac{1}{3} \left[ \frac{\partial R}{\partial Z} \right]^2 dZ
\le 0 .
\label{eqn:RHS}
\end{align}
The left term in Eq. (\ref{eqn:hyperbolic_form}) can also be expressed as
\begin{align}
\notag
\int_{Z_1}^{Z_2} R \frac{\partial G}{\partial T}dZ
&= \int_{Z_1}^{Z_2} (G^{3/2} - G^{3/2}_S) \frac{\partial G}{\partial T}dZ
\\\notag
&= \int_{Z_1}^{Z_2} \frac{\partial}{\partial T}\left(\frac{2}{5}G^{5/2} - G G^{3/2}_S\right)dZ
\\
&= \frac{\partial}{\partial T} \int_{Z_1}^{Z_2} \left(\frac{2}{5} G^{5/2} - G^{3/2}_S G + \frac{3}{5} G^{5/2}_S \right) dZ
\label{eqn:freeenergydensity}
\end{align}
where the last term in Eq. (\ref{eqn:freeenergydensity}) has been added, without loss of generality, to ensure the integrand vanishes for $G=G_S$, which essentially sets the value of the ground state energy. Given that $\delta \mathfrak{F}/\delta G = R[G(Z)]$, it is evident that the integral defined in Eq. (\ref{eqn:freeenergydensity}) is none other than the Lyapunov free energy, which can be re-expressed in terms of the interface function $H$:
\begin{equation}
\mathfrak{F}(H) = \int_{Z_1}^{Z_2} \left(\frac{2}{5} H^{5} - H^{3}_S H^2 + \frac{3}{5} H^{5}_S \right)dZ .
\label{eqn:F(H)}
\end{equation}
Equating Eq. (\ref{eqn:RHS}) and Eq. (\ref{eqn:freeenergydensity}) then yields the relation
\begin{equation}
\frac{\partial \mathfrak{F}(H)}{\partial T}
= - \frac{1}{3}
\int_{Z_1}^{Z_2} \left[ \frac{\partial}{\partial Z}(H^3 - H^3_S) \right]^2 dZ  \le 0 ,
\label{eqn:dFdT}
\end{equation}
where the final equality holds only for initial states $H(Z,T)$ exactly equal to the stationary state solutions $H_S(Z)$. This finding serves as proof that the flow described is asymptotically stable in the Lyapunov sense i.e. arbitrary initial distributions $H(Z,T)$ decay in time toward the stationary solution $H_S$, characterized by a vanishing first variation and positive second variation, namely
\begin{align}
\left. \delta \mathfrak{F}(G)\right |_{G_S}
& = \int^{Z_1}_{Z_2} \left. \frac{\delta \mathfrak{F}}{\delta G} \right|_{G_S} \delta G \, dZ \\
& = \int^{Z_1}_{Z_2} \left. R(G)\right|_{G_S} \, \delta G \, dZ = 0 \, \, \, \textrm{and} \\
\left. \delta^2 \mathfrak{F}(G)\right|_{G_S}
& = \int^{Z_1}_{Z_2} \left. \frac{\delta^2 \mathfrak{F}}{\delta G^2}\right|_{G_S} (\delta G)^2 \, dZ\\
& = \int^{Z_1}_{Z_2} \frac{3}{2}\, G^{1/2}_S \,(\delta G)^2 \, dZ > 0 .
\end{align}

To complete this part of the proof, we show that the stable stationary states $H_S$ are in fact also unique. Suppose then that in addition to the already specified steady state solution $H_S$ there exists another steady solution $H_{S2}$ satisfying the same boundary conditions. Since $H_{S2}$ is time independent, it must satisfy the relation $\partial \mathfrak{F}(H_{S2})/\partial T = -(1/3)\int_{Z_1}^{Z_2} \left[ \partial (H_{S2}^3 - H_S^3)/\partial Z \right]^2 dZ = 0$. Clearly this is only possible if $H_{S2} = (H^3_S + C_o)^{1/3}$, where the constant $C_o$ must equal zero in order for $H_{S2}$ to satisfy the same boundary conditions as $H_{S2}$. This result therefore establishes that the solution $H_S$ is indeed unique.

The proof above requires that both the stationary $H_S(Z)$ and transient $H(Z,T)$ solutions be everywhere strictly positive. For solutions satisfying Dirichlet boundary conditions, these requirements are easily met. Consider any positive initial state $H(Z,T=0)$ which redistributes its height in time according to Eq. (\ref{eqn:InterfaceEqn}). Were there a point within the domain where $H(Z,T) \to 0$, then the local interface would have to satisfy $\partial H/\partial T <0$ and the local curvature $\partial^2 H/\partial Z^2$ would have to become sufficiently negative to satisfy the balance of terms Eq. (\ref{eqn:InterfaceEqn}). The local interface would then have to develop a local protrusion (negative curvature) and not a local dimple leading to rupture (positive curvature), in contradiction to the assumption of positive curvature required at that local minimum. Therefore, all positive initial states remain positive in time. For Dirichlet conditions then, stationary solutions $H_S(Z)$ are \textit{globally} exponentially stable. For the remainder stationary solutions subject to a Neumann boundary condition, it may become the case that too high a flux condition leads to one of more points where the local film thickness vanishes. In the vicinity of such points, the local interface slope will become increasingly large and eventually violate the slender limit approximation. Perhaps more importantly, disjoining pressure effects, known to modify the flow in ultrathin regions of a liquid film, must then be incorporated into the model interface equation \cite{TS87,SP01}. In such cases then, the proof above only establishes asymptotic stability, not global asymptotic stability.

Next, we demonstrate an even stronger statement regarding the asymptotic stability of stationary states, namely that the stationary solutions $H_S$ represent \textit{exponentially stable} equilibria of Eq. (\ref{eqn:H2Master}). A useful relation for the function $R(Z)$ defined in Eq. (\ref{eqn:R(Z)}) can be obtained by first applying the Poincar\'e-Friedrichs inequality in one dimension \cite{SB16} according to which
\begin{equation}
\label{thm:Poincare_lemma}
\int_{Z_1}^{Z_2} \left (\frac{dR}{dZ} \right)^2 \, dZ \ge \frac{1}{(Z_2 - Z_1)^2} \int_{Z_1}^{Z_2} R^2 (Z) dZ
\end{equation}
subject to the constraint that there is some interior point $Z^*$ in the bounded domain $Z_1 \le Z^* \le Z_2$ where $R(Z^*) = 0$. This must always be the case, however, since $H(Z,T)$ must satisfy the same boundary conditions as $H_S$. Given that $H(Z,T)$ and $H_S(Z)$ are strictly positive throughout the domain, it is possible to construct an upper bound on $\partial \mathfrak{F}/\partial T$ in Eq. (\ref{eqn:dFdT}) as follows:

\begin{align}
\notag
\frac{\partial \mathfrak{F}(H)}{\partial T}
& = -\frac{1}{3} \int_{Z_1}^{Z_2} \left[ \frac{\partial}{\partial Z}(H^3 - H^3_S) \right]^2 dZ
\\
\notag
& \le -\frac{1}{3(Z_2-Z_1)^2} \int_{Z_1}^{Z_2} (H^3 - H^3_S)^2 dZ
\\
\notag
& < - \frac{1}{3 \, \mathcal{C} \, (Z_2-Z_1)^2} \int_{Z_1}^{Z_2} (H^3 - H^3_S)^2 \,\, \times
\\
\notag
& \left \{\frac{2 H^3 + 4 H^2 H_S + 6 H H^2_S + 3 H^3_S}{5(H^2 + H H_S + H^2_S)^2} \right \} dZ
\\
\notag
& =  - \frac{1}{3 \,\mathcal{C}\, (Z_2-Z_1)^2} \int_{Z_1}^{Z_2} \left(\frac{2}{5} H^5 - H^3_S H^2 + \frac{3}{5} H^5_S \right) dZ
\\
\notag
& =  - \frac{\mathfrak{F}(H)}{3\, \mathcal{C} \,(Z_2-Z_1)^2}  ,
\end{align}
where $\mathcal{C}$, which is strictly positive, is defined as
\begin{equation}
\label{eqn:constC}
\mathcal{C} = \min_{T,Z} \!
\left \{\frac{2 H^3 + 4 H^2 H_S + 6 H H^2_S+ 3 H^3_S}{5(H^2 + H H_S + H^2_S)^2} \right \} .
\end{equation}
(See Appendix \ref{sec:A1} for further discussion regarding the boundedness property of $\mathcal{C}$). The final inequality sought is then given by
\begin{equation}
\frac{\mathfrak{F}(H,T)}{\mathfrak{F}(H,T=0)} <
\exp \left\{ \frac{-\,T}{3 \mathcal{C} (Z_2 - Z_1)^2} \right \} ,
\label{eqn:Fexponential}
\end{equation}
which confirms that the Lyapunov free energy of any disturbed state $H(Z,T)$ satisfying the same boundary conditions as the initial stationary state will decay back to the stationary state $H_S$ at least as fast as an exponential with a decay rate that scales with the square of the spatial domain size. This proof applies to all categories of stationary solutions discussed in Section \ref{StatSolns}. For disturbance functions subject either to two Dirichlet conditions or one Dirichlet condition and one Neumann (constant flux) or constant volume condition, the proof is trivial since either $R(Z_1)$ or $R(Z_2)$ vanishes identically. For solutions $H_S$ that correspond to fixed flux $Q_S$ at one boundary and constant volume $V_S$, the proof above also applies since there always exists an interior point $Z_1 \le Z^* \le Z_2$ where $R\left(Z^*\right) = 0$. This follows because two solutions $H$ and $H_S$ cannot have the same constant volume unless the functions $H(Z,T)$ and $H_S(Z)$ undergo at least one crossing point within the domain.

\section{Stability of self-similar solutions}
\label{sec:selfsimilar}
We next seek unperturbed (i.e. base state) transient solutions to Eq. (\ref{eqn:H2Master}) for non-conserved volume which manifest self-similarity. A global stability argument as presented in Section \ref{sec:nonlinstab} for stationary solutions proved unsuccessful since the self-similar form of the governing interface equation cannot be converted into gradient flow form. Therefore we instead applied a generalized linear stability analysis \cite{FI96} which helps determine whether infinitesimal non-modal disturbances undergo any transient or asymptotic amplification.

\subsection{Self-similar solutions with time independent Dirichlet conditions}
\label{sec:SelfSimDirichlet}
Previous studies have delineated the conditions leading to existence and uniqueness of self-similar solutions as well as the attraction of spatially confined initial distributions toward self-similar base states \cite{V07}. Here we focus on volume non-conserving positive states $S(\eta)$ consistent with time-independent Dirichlet boundary conditions imposed at the domain endpoints, namely $S(0)=1$ and $S(\eta \rightarrow \eta_B)= \textrm{const}$ where $\eta_B$ denotes a location far downstream of the origin. The Dirichlet condition at the origin can be set to unity without loss of generality since as evident from Eq. (\ref{eqn:SelfSimEqn}), a rescaling involving a multiplicative factor of $S(0)$ leaves the governing equation unchanged.

In general, for self-similarity to hold, there can be no intrinsic length or time scale imposed on the flow, in contrast to the steady state solutions examined in the previous section which depend on the groove length $Z_2 - Z_1$ . A simple scaling analysis of Eq. (\ref{eqn:H2Master}) reveals that self-similar solutions may be possible whenever $T << L^2/H \sim O(L/\varepsilon)$. To find such solutions, it is convenient to expand and rewrite Eq. (\ref{eqn:H2Master}) in the form
\begin{equation}
\frac{\partial H}{\partial T} - \frac{H}{2} \frac{\partial^2 H}{\partial Z^2} - \left (\frac{\partial H }{\partial Z}\right)^2 = 0 .
\label{eqn:InterfaceEqn}
\end{equation}
The ansatz $H_{\textrm{sim}}(\eta, T)$ defined by
\begin{align}
H_{\textrm{sim}}(\eta) &= T^{2\beta - 1} S(\eta) \,\,\,\, \textrm{where}\\
\eta  & = \frac{Z}{T^\beta} \,\,\, \textrm{for} \,\, \beta>0,
\end{align}
allows for a large class of self-similar solutions \cite{V07,WL98} satisfying the general second order nonlinear differential equation
\begin{equation}
\frac{S}{2} S_{\eta \eta} + (S_\eta)^2 + \beta \eta S_\eta + (1-2\beta)S = 0 .
\label{eqn:SPDE}
\end{equation}
Inspection of the asymptotic behavior of $S(\eta)$ as $\eta \to \infty$ helps ascertain
what range of exponents $\beta$ are required for bounded non-terminating (i.e. $S>0$) states such that $S_{\eta \eta}$ and $S_\eta$ asymptotically approach zero as $\eta \to \infty$ . While the first two terms on the left side of Eq. (\ref{eqn:SPDE}) then vanish identically, care must be taken with regard to the third term which couples an increasingly large value of $\eta$ with a diminishingly small term $S_{\eta}$. Balancing  the third and fourth terms yields the proper asymptotic scaling, namely $dS/S \sim [(2\beta -1)/\beta] d\eta/\eta$, and hence $S(\eta \to \infty) \sim \eta^{(2\beta -1)/\beta}$. Therefore, only the range $0 < \beta \leq 1/2$ yields bounded non-terminating self similar states.

Boundary conditions also impose constraints on the allowable values of the exponent $\beta$. For example, enforcement of constant liquid volume $V \simeq \int_{Z_1}^{Z_2} H^2 dZ = T^{5\beta-2} \int_{Z_1}^{Z_2} S^2 d\eta$ is only consistent with $\beta = 2/5$. According to Eq. (\ref{eqn:QS}), enforcement of a constant flux boundary condition $Q = - \widehat{A}(\alpha/\theta) H^2 (\partial H/\partial Z) = \textsf{const}= - \widehat{A}(\alpha/\theta)T^{3\beta - 2} S^2 S_\eta$ is only consistent with $\beta = 3/5$. Clearly then, a constant flux boundary condition (Neumann condition) is therefore inconsistent with bounded solutions.

In what follows, we restrict attention to the value $\beta = 1/2$, which accords with the Washburn relation and allows enforcement of time-independent Dirichlet boundary conditions. For this category of solutions, the non-dimensional flux defined in Eq. (\ref{eqn:NondimFlux}) is represented by
\begin{equation}
  Q(\eta,T) = - \frac{\widehat{A}(\alpha, \theta)}{T^{1/2}} S^2 \frac{\partial S}{\partial \eta},
\label{eqn:NondimSelfSimFlux}
\end{equation}
The self-similar solution $S(\eta)$ then satisfies the equation \cite{WDec01}:
\begin{equation}
S S_{\eta \eta} + \eta S_\eta + 2 (S_\eta)^2 = 0 .
\label{eqn:SelfSimEqn}
\end{equation}

\begin{figure}
\centering
\includegraphics[width = 8.0 cm]{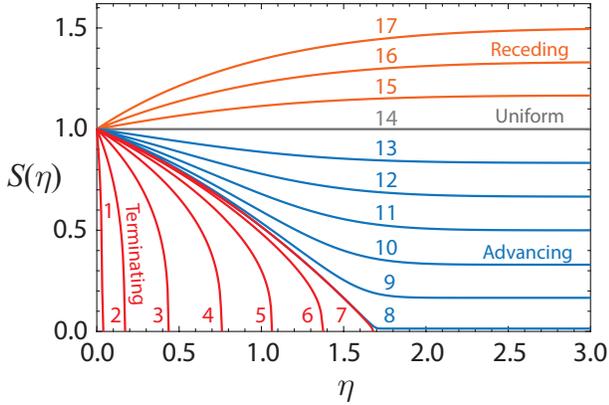}
\caption{(Color online) Representative self-similar solutions $S(\eta)$ for terminating (1-7), advancing (8-13), uniform (14) and receding (15-17) states. The computational domain used in numerically solving for these solutions was $[0 \leq \eta \leq 80]$. (Only  the range $[0 \leq \eta \leq 3.0]$ is shown in the figure since the downstream behavior remains essentially unchanged beyond that value.) All solutions satisfy the Dirichlet condition $S(\eta= 0) = 1$ at the origin. Solutions 1-7 exhibit interface slopes at the origin given by $S_\eta(0)$ =-10, -2, -0.8, -0.5, -0.4, -0.36, and -0.3492. Solutions  8-13 satisfy far field Dirichlet conditions given respectively by $S(80)$ = 0.01, 0.17, 0.33, 0.5, 0.67 and 0.83. (Resulting values for the interface slopes at the origin are $S_\eta(0)$ = -0.3491, -0.3418, -0.3185, -0.2745, -0.2090, and -0.1185, respectively.) The line denoted by 14 represents a uniform solution where $S(\eta) = 1.0$. Solutions 15-17 satisfy far field Dirichlet conditions given respectively by $S(80)$ = 1.17, 1.33 and 1.5. (Resulting values for the interface slopes at the origin are $S_\eta(0)$ = 0.1485, 0.3283 and 0.5451, respectively.)}
\label{fig:selfsimilar_solutions}
\end{figure}

To ascertain the interface shape of these solutions, we numerically solved Eq. (\ref{eqn:SelfSimEqn}) by rewriting the second order equation as a system of first order equations and using the \textsf{ODE45} solver in Matlab \cite{Matlab2015a}. Shown in Fig. \ref{fig:selfsimilar_solutions} are representative solutions for receding, uniform, advancing and terminating states  $S(\eta)$ satisfying the far field Dirichlet conditions shown. According to Eq. (\ref{eqn:NondimSelfSimFlux}), solutions with $S_\eta > 0$ correspond to states with net liquid flux to the left, designated \textit{receding states}, while solutions with $S_\eta < 0$ correspond to a net liquid flux to the right, designated \textit{advancing} states. The solution for which $S\eta$ vanishes everywhere, which corresponds to a zero flux solution in self-similar coordinates, is designed a \textit{uniform} state. It represents an exception in that it is the only solution which satisfies volume conservation. Solutions whose advancing front are characterized by a vanishing value of $S(\eta)$ are likewise designated \textit{terminating} states. The numerical solutions indicate that solutions undergo termination only when the interface slope at the origin $S_\eta(0) \gtrsim 0.349$.

\subsection{Generalized non-modal linear stability of interface self-similar solutions }
\label{sec:genlinstab_background}
It is by now well understood that traditional modal stability analysis, wherein  instability derives from exponentially growing normal modes of the governing linearized autonomous disturbance operator, is an inappropriate investigatory tool for examining transient stability of systems governed by non-normal operators \cite{FI96}. This is because the modal spectrum, and in particular the eigenvalue with maximum real part, properly captures the system response to infinitesimal disturbances at all times only for linearized disturbance operators which are normal; for disturbance operators which are non-normal, it describes the response only at infinite time. The case of non-normal operators therefore requires implementation of a generalized (non-modal) or transient growth stability analysis, as discussed in Refs. [\onlinecite{TD93,FI96}] and references therein. The important distinction is that while for normal operators instability arises from a single, fastest growing, exponentially unstable eigenmode which dominates at all times, instability in systems governed by non-normal operators derives from non-modal growth of two or more interacting nonorthogonal eigenmodes at finite times while only asymptotically resolving to the disturbance function characterized by the eigenvalue with maximum positive real part.

Equations of motion which give rise to inhomogeneous base states often yield linearized  operators which are non-normal. Such non-normal operators commonly arise in many thin film liquid systems \cite{DT03b,DT03c}. Given that the base state solutions to Eq. (\ref{eqn:SelfSimEqn}) include advancing and receding states which vary with the self-similar variable, we therefore appeal to a transient stability analysis to investigate their behavior. The stability of terminating solutions shown in Fig. \ref{fig:selfsimilar_solutions}, however, will not be examined in this work due to the fact that the point of termination requires that additional terms be included in the governing equation of motion to relieve the diverging stress known to occur at a moving contact line \cite{DT03a,DT04,DT06}.

To begin, we linearize the general solution according to
\begin{equation}
H(\eta, \tau) = S(\eta) + \delta G(\eta, \tau),
\label{eqn:disturbform}
\end{equation}
where $S(\eta)$ satisfies Eq. (\ref{eqn:SelfSimEqn}) with the given boundary conditions, $G(\eta,\tau)$ represents the non-modal disturbance function and $\delta \ll 1$ denotes the small expansion parameter for linearization. Substituting this form into Eq. (\ref{eqn:InterfaceEqn}) yields the governing linear disturbance equation to order $\delta$, namely:
\begin{equation}
\frac{\partial G}{\partial \mathrm{\tau}} = \mathcal{L} [G],
\label{eqn:operatorL}
\end{equation}
where
\begin{align}
\notag
\mathcal{L} & = \, \frac{S}{2}\frac{\partial^2}{\partial \eta^2} +
\left (2 \frac{d S}{d \eta} + \frac{\eta}{2} \right) \frac{\partial}{\partial \eta} + \frac{1}{2}\frac{d S}{d \eta^2}, \\
\mathcal{L}^\dag & = \, \frac{S}{2}\frac{\partial^2}{\partial \eta^2} -
\left (\frac{\eta}{2} + \frac{d S}{d \eta} \right) \frac{\partial}{\partial \eta} - \left(\frac{1}{2}+\frac{d^2 S}{d \eta^2}\right) .
\label{eqn:LDefinitions}
\end{align}
Here, $\mathcal{L}^\dag$ denotes the adjoint of the linear autonomous operator $\mathcal{L}$, $\tau = \ln T$ and $G(\eta,\tau) = \exp(\mathcal{L} \tau) G(\eta,0)$. Under the $L_2$ (Euclidean) norm, the adjoint $\mathcal{L}^\dag$ represents the unique linear operator where $\int_0^{\eta_B} v(\eta) \mathcal{L} \left\{w(\eta)\right\} d\eta = \int_0^{\eta_B}\mathcal{L}^\dag \left\{v(\eta)\right\} w(\eta) d\eta$ for all sufficiently smooth functions $v$ and $w$ which are $L_2$-integrable and satisfy homogeneous Dirichlet boundary conditions at the boundary points $(0,{\eta_B})$. For all non-uniform solutions $S$, $\mathcal{L} \mathcal{L}^\dag \neq \mathcal{L}^\dag \mathcal{L}$ so that $\mathcal{L}$ is non-normal. The transient amplification of disturbances $G(\eta, \tau)$ cannot therefore simply be ascertained from the eigenspectrum of $\mathcal{L}$. Disturbance amplification over an interval in time $\tau$ will instead be quantified by the function
\begin{align}
\sigma & = \frac{\|G(\eta,\tau)\|}{\|G(\eta,0)\|} = \, \frac{\left<G(\eta,\tau)|G(\eta,\tau)\right>^{1/2}}{\left< G(\eta,0)|G(\eta,0)\right>^{1/2}} \\
& = \frac{\left<\exp(\mathcal{L}^\dag \tau) \exp(\mathcal{L} \tau)G(\eta,0)|G(\eta,0) \right>^{1/2}}{\left< G(\eta,0)|G(\eta,0)\right>^{1/2}} &\\
& \leq \, \|\exp(\mathcal{L} \tau)\|,
\label{eqn:amplifsigma}
\end{align}
where the notation $\| \cdot \|^2$ represents the Euclidean norm $\left\langle\cdot\,|\,\cdot\right\rangle^{1/2}$ when applied to functions and the operator norm when applied to operators, as in Eq. (\ref{eqn:amplifsigma}).

To gain further insight into the growth or decay of disturbances at finite times, it is useful to consider the discretized operator $\exp(\mathcal{L} \tau)$ in diagonalized form such that
\begin{equation}
\exp(\mathcal{L} \tau)= \mathcal{S} \exp(\Lambda \tau) \mathcal{S}^{-1}
\end{equation}
where $\exp(\Lambda \tau)$ denotes a diagonal matrix whose entries are the eigenvalues of $\mathcal{L}$ arranged in decreasing order and $\mathcal{S}$ represents the matrix whose columns are the corresponding eigenfunctions. It can then be shown \cite{FI96} that
\begin{align}
\exp(\beta_\textrm{max} \tau) & \leq  \|\exp(\mathcal{L} \tau)\|
= \|\mathcal{S} \exp(\Lambda \tau) \mathcal{S}^{-1}\| \\
& \leq \|\mathcal{S}\|\|\mathcal{S}\|^{-1}\exp(\beta_\textrm{max} \tau),
\label{eqn:inequal}
\end{align}
where $\beta_{\textrm{max}}$ is the eigenvalue of $\mathcal{L}$ with maximum real part, also known as the spectral abscissa of $\mathcal{L}$. A matrix operator is normal if and only if it is unitarily diagonalizable such that $\|S\|\|S\|^{-1} = 1$, in which case the maximum amplification at time $\tau$ over all initial conditions $G(\eta,0)$ is given by $\exp(\beta_{\textrm{max}} \tau)$. For non-normal operators, however, $\|S\|\|S\|^{-1} \geq 1$ and so the maximum disturbance amplification at finite time $\tau$ can exceed the asymptotic value $\exp(\beta_\textrm{max} \tau)$, sometimes significantly so. We note too from Eq. (\ref{eqn:amplifsigma}) that as $\tau \rightarrow \infty$, the quantity $d \ln(\|\exp(\mathcal{L} \tau)\|)/d \tau \rightarrow \beta_\textrm{max}$. While a finding that $\beta_{\textrm{max}} < 0$ establishes that a system is \textit{asymptotically} linearly stable, it does not preclude the possibility of large positive transient growth (i.e. linear instability) at finite time.

We now determine the upper bound on the instantaneous rate of disturbance growth at time $\tau$ given by
\begin{align}
\omega & = \, \frac{1}{\sigma} \frac{\partial \sigma}{\partial \tau} = \frac{\partial \ln \sigma}{\partial \tau}\\
& = \frac{1}{\|G(\eta,\tau)\|}\frac{\partial}{\partial \tau} \left\langle G(\eta,\tau)|G(\eta,\tau)\right\rangle^{1/2} &\\
& = \,  \frac{\left\langle\mathcal{L} G(\eta,\tau)|G(\eta,\tau)\right\rangle + \left\langle G(\eta,\tau)|\mathcal{L} G(\eta,\tau)\right\rangle}{2\|G(\eta,\tau)\|^2}\\
& = \, \frac{1}{\|G(\eta,\tau)\|^2} \left \langle\frac{\mathcal{L} + \mathcal{L}^\dag}{2}\ G(\eta,\tau)| G(\eta,\tau) \right\rangle&\\
& \leq \,  \textrm{max} \, \left\{\lambda \left(\frac{\mathcal{L} + \mathcal{L}^\dag}{2}\right) \right \}
\label{eqn:omegamax}\\
& \equiv \omega_{\textrm{max}}.
\label{eqn:sigma_exp_growth}
\end{align}
The quantity $\omega_{\textrm{max}}$, also known as the numerical abscissa, represents  the maximum eigenvalue of the operator sum, $(\mathcal{L} + \mathcal{L}^\dag)/2$. Since this combined operator is self-adjoint, its eigenvalues are strictly real. Therefore, while $\beta_{\textrm{max}}$---the eigenvalue of $\mathcal{L}$ with maximum real part---governs the asymptotic stability as $\tau \rightarrow \infty$, it is the value $\textrm{max} \, \{\lambda\}$ which governs the transient stability of the linearized system at finite times. Since physical systems undergo flow instabilities at finite time, it is this latter value which is of most interest. According to this definition then, if it can be shown that $\omega_{\textrm{max}} < 0 $ for all times $\tau$, then instantaneous disturbance growth will always be suppressed and the system can be deemed linearly stable to infinitesimal perturbations. Below we establish the generalized linear stability of self-similar solutions $S(\eta) > 0$ subject to time-independent Dirichlet conditions for volume non-conserving base states describing inertia-less flow in a slender open triangular channel.  We first compute analytic bounds on instantaneous disturbance growth for advancing and receding films followed by results obtained from direct numerical computation.

Having introduced the main concepts underlying transient growth analysis, we conclude this section with a few words about the stability of uniform stationary solutions to Eq. (\ref{eqn:InterfaceEqn}). Weislogel \cite{W01} previously conducted a modal linear stability analysis for an infinitely long, slender, quiescent liquid filament of uniform thickness $d$ confined to the interior corner of an open triangular channel. He showed that axial sinusoidal perturbations of any wavelength $\ell$ all undergo rapid decay with a dimensional time constant proportional to $\mu \ell^2 /\gamma d$. In our study, the governing dimensionless linear operator corresponding to a stationary uniform state is given by $\mathcal{L} = (d/2) \, \partial^2/\partial Z^2$, an operator which is self-adjoint and therefore normal, whose eigenvalues are strictly real, positive and given by $(d/2)(n \pi /L)^2$ for $n=1$, 2, etc. According to Eq. \ref{eqn:operatorL} (when expressed in the original laboratory frame coordinates $(Z,T))$, the least stable disturbance $\delta G (Z,T)$ will evolve in time according to $\exp[-(d/2)(\pi /L)^2 T]$ and therefore decay away, restoring the system back to the initial uniform state. This observation confirms Weislogel's result that uniform quiescent liquid filaments confined to a slender open triangular channel are linearly stable to arbitrary infinitesimal disturbances of any wavenumber. This result holds for any of the four sets of boundary conditions discussed earlier, namely two Dirichlet conditions, one Dirichlet and one Neumann condition, one Dirichlet condition and constant volume or one Neumann condition and constant volume.

\subsection{Generalized stability of volume non-conserving self-similar solutions}
\label{sec:selfsimilar_stability}
In what follows, we examine the generalized linear stability of self-similar non-terminating states on the finite domain $0 \leq \eta \leq \eta_B$. To proceed, we first note that the linear autonomous operator $(\mathcal{L} + \mathcal{L}^\dag)/2$ defined in Eq. (\ref{eqn:LDefinitions}) (which is also self-adjoint and therefore normal) and given by
\begin{align}
\frac{\mathcal{L} + \mathcal{L}^\dag}{2} = & \, \frac{\partial}{\partial \eta} \left(\frac{S(\eta)}{2} \frac{\partial}{\partial \eta}\right) - \frac{1}{4}\left( 1 + \frac{\partial^2S}{\partial \eta^2} \right)
\label{eqn:L}
\\
=& \, \frac{S}{2}\frac{\partial^2}{\partial \eta^2} + \frac{1}{2}\frac{\partial S}{\partial \eta}\frac{\partial}{\partial \eta} - \frac{1}{4}\left( 1 + \frac{\partial^2S}{\partial \eta^2} \right) ,
\label{eqn:Ldag}
\end{align}
satisfies a Sturm-Liouville eigenvalue equation where
\begin{equation}
\left(\frac{\mathcal{L} + \mathcal{L}^\dag}{2}\right ) \, G(\eta,\tau) = - \lambda \, G(\eta,\tau) .
\label{eqn:SLtype}
\end{equation}
provided $S(\eta)>0$ and $S$, $dS/d\eta$ and $d^2 S/dS^2$ are continuous within the interval $0 \leq \eta \leq \infty$. As a result, the eigenvalues $\lambda$ are strictly real and the corresponding eigenfunctions form a complete orthogonal set of basis functions. According to the definitions in Eq. (\ref{eqn:sigma_exp_growth}) then, if it can be shown that all the eigenvalues of $\mathcal{L} + \mathcal{L}^\dag/2$ are strictly positive, then $\omega_{max} < 0$ and disturbance growth is suppressed.

Before proceeding with the behavior of advancing or receding states, we first note that the linear stability of the uniform state $S(\eta) = 1$ is easy to confirm. From the definitions given by Eqs. (\ref{eqn:Ldag}) and (\ref{eqn:SLtype}), the relevant operator sum reduces to $(1/2) \, (\partial^2/\partial \eta^2) - (1/4)$, which is self-adjoint and therefore normal and whose eigenvalues are strictly real, positive and given by $(1/2)(n \pi /L)^2 + 1/4$ for $n=1$, 2, etc. According to Eq. (\ref{eqn:SLtype}), the least stable disturbance $\delta G(\eta,\tau)$ will evolve in time according to $\exp{-(1/2)[(n \pi /L)^2 + 1/4T]\, \tau}$ and therefore decay away, restoring the system back to the initial uniform self-similar state.

It is often the case that the Rayleigh quotient \cite{H04} can be used to establish bounds on growth or decay by helping determine the overall sign of eigenvalues corresponding to Sturm-Liouville operators. In our case, however, this approach yields no useful information. We therefore instead appeal to the following lemma:
\begin{lemma}
Given a second order Sturm-Liouville eigenvalue equation of the form $(P\, G_\eta)_\eta
+ Q\,G = - \lambda G$ where $P > 0$, $G \neq 0$, $\lambda$ is real, and $G(0) = G(\eta_B) = 0$, then $\lambda_{\textrm{min}} >(-Q)_{\textrm{min}}$,
\label{thm:SL_lemma}
\end{lemma}
where subscripts denote differentiation with respect to the domain coordinate $\eta$. In order for $G$ to satisfy homogeneous conditions at the boundary points, it must be the case that there exists an extremum within the domain $[0,\eta_B]$ where $G_\eta(\eta^*)=0$. Without loss of generality, the first such extremum from the origin may be assumed to be a local maximum in $G$ for $G > 0$. (Note that $G$ can always be made positive at this point by multiplying the local value by -1 and still remain an eigenfunction). Because $P G_\eta$ must change sign on either side of this maximum, then at some point in its vicinity, $(P G_\eta)_\eta < 0$. This yields the relation $\lambda > -Q(\eta^*)$, which therefore implies $\lambda_{\min} > -Q(\eta^*) \ge (-Q)_{\min}$, where $(-Q)_{\min}$ denotes the smallest value of $-Q$ within the domain.

\subsubsection{Transient and asymptotic stability of advancing solutions $S(\eta)$}
We first examine the stability of advancing self-similar solutions shown in Fig. \ref{fig:selfsimilar_solutions}. These solutions are all characterized by a downstream boundary value smaller than the inlet value, namely $S(\eta_B) = \textrm{const} < S(0) = 1$, and an inlet slope which is always negative, namely $S_{\eta}(0) < 0 $. Here and in what follows, $\eta_B$ represents the coordinate where the downstream boundary condition is applied.

Establishing that such states are linearly stable requires a finding that the smallest eigenvalue of $(\mathcal{L} + \mathcal{L}^\dag)/2$, namely $\lambda_{\textrm{min}}$, is positive, which then requires
\begin{equation}
\lambda_{\textrm{min}} > (-Q)_{\textrm{min}} = (1 + S^{\textrm{min}}_{\eta \eta})/4 > 0 ,
\label{eqn:Qminadv}
\end{equation}
which in turn requires that $S^{\textrm{min}}_{\eta \eta} > -1$. This minimum value can occur either at some interior point $\eta_{\textsf{int}}$ or at the boundary points $\eta = 0, \eta_B$, which shall be examined separately. When the minimum value $S^{\textrm{min}}_{\eta \eta}$ occurs at the downstream boundary point $\eta_B$, the inequality is easily satisfied for all solutions $S$, whether advancing or receding, since as evident from Fig. \ref{fig:selfsimilar_solutions}, all the self-similar solutions asymptote to a uniform thickness, which therefore yields values $S_{\eta \eta}(\eta \rightarrow \eta_B) \to 0 > -1$.

When $S^{\textrm{min}}_{\eta \eta}$ occurs at an interior point $\eta_{\textsf{int}}$, it will then be the case that $S^{\textrm{min}}_{\eta \eta \eta}(\eta_{\textsf{int}})=0$. Differentiation of Eq. (\ref{eqn:SelfSimEqn}) then yields the corresponding third order  equation evaluated at the point $\eta_{\textsf{int}}$
\begin{equation}
\bigg[ \left(\frac{S_\eta}{S^2}\right) \big[- S + \left( \eta + 2 S_\eta \right) \left( \eta + 5 S_\eta \right) \big] \bigg] \left(\eta_{\textsf{int}} \right)=0
\end{equation}
whose solutions are given by
\begin{align}
S_\eta (\eta_{\textsf{int}}) =& \, 0 \, \, \, \, \, \textrm{or} \\
= & \,\frac{-7 \eta_{\textsf{int}} \pm \sqrt{9 \eta^2_{\textsf{int}} + 40 S(\eta_{\textsf{int}})}}{20} ,
\end{align}
which when substituted into Eq. (\ref{eqn:SelfSimEqn}) provides the value of the curvature at that interior point, namely
\begin{align}
S_{\eta \eta}(\eta_{\textsf{int}}) &= \, 0  \, \, \, \, \, \textrm{or} \\
&= \, -\frac{1}{5} + \frac{3\eta^2_{\textsf{int}} \pm \eta_{\textsf{int}}
\sqrt{9 \eta^2_{\textsf{int}} + 40 S(\eta_{\textsf{int}})}}{50 S(\eta_{\textsf{int}})} .
\end{align}
The minimal value of this relation is attained for the negative root, which as shown is always less than -1:
\begin{align}
\notag
&S^{\textrm{min}}_{\eta \eta}(\eta_{\textsf{int}}) = -\frac{1}{5} + \frac{3\eta_{\textsf{int}}^2 - \eta_{\textsf{int}}\sqrt{9 \eta_{\textsf{int}}^2 + 40 S(\eta_{\textsf{int}})}}{50 S(\eta_{\textsf{int}})} \\
\notag
&=-\frac{1}{5} + \frac{3}{50}~\frac{\eta_{\textsf{int}}}{\sqrt{S(\eta_{\textsf{int}})}}~ \left( \sqrt{\frac{\eta_{\textsf{int}}^2}{S(\eta_{\textsf{int}})}} - \sqrt{\frac{\eta_{\textsf{int}}^2}{S(\eta_{\textsf{int}})} + \frac{40}{9}} \right) \\
&\geq -\frac{1}{5} + \frac{3}{50}~ \times \left(-\frac{20}{9}\right)\\
& > -1 .
\end{align}
The last inequality derives from the general relation $\eta (\sqrt{\eta^2}-\sqrt{\eta^2+k}) \ge -k/2$ for $\eta$ and $k$ real and positive.

When the minimum value of $S_{\eta \eta}$ occurs at the origin, evaluation of Eq. (\ref{eqn:SelfSimEqn}) subject to the Dirichlet condition $S(0)=1$ yields the criterion for stability, namely $S_{\eta \eta}(0) = -2 S_\eta^2 (0) > -1$. Since advancing solutions always exhibit negative slopes at the origin, this criterion can be rewritten as $S_\eta (0) < -1/\sqrt{2}$. Our numerical results show that in order for advancing solutions to remain strictly positive throughout the domain $[0,L\eta_B]$ and not yield a termination point, it must be the case that $S_\eta (0) \lesssim -0.349$, which is clearly greater than $-1/\sqrt{2}$. The following geometric argument supports this conclusion as well.

\begin{figure}
\centering
\includegraphics[scale=1.0]{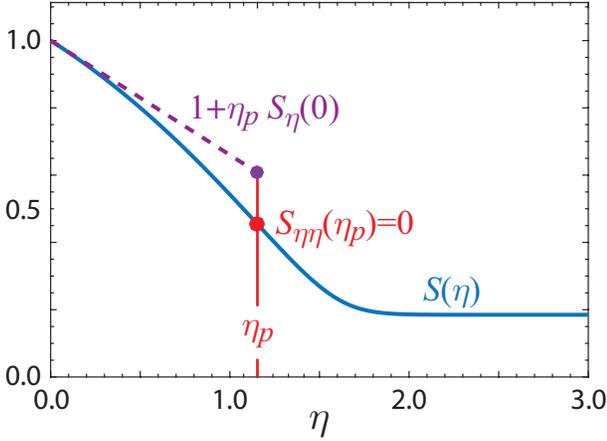}
\caption{(Color online) Representative self-similar solution $S(\eta)$ (blue solid line) for advancing state. The (red) dot designates the point $\eta_P$ where there occurs the first inflection point away from the origin where  $S_{\eta \eta}(\eta_p) = 0$. The (purple) dashed line represents a linear extension of the Taylor expansion of $S(\eta)$ about the origin to the point $\eta_P$.}
\label{fig:Inflection}
\end{figure}

Since $S_{\eta \eta}(0) = -2 S_\eta^2 (0) < 0$ while $S_{\eta \eta}(\eta_B) = 0$, there must occur at least one inflection point in the domain $[0,\eta_B]$. Consider the first such inflection point at position $\eta_p$ where $S_{\eta \eta}(\eta_p) = 0$. As shown in Fig. \ref{fig:Inflection}, since $S(\eta_p)$ lies below the extended line $1+\eta_p S_\eta(0)$, then $S(\eta_p) < 1+ \eta_p S_\eta(0) = 1 - 2 S_\eta (\eta_p) S_\eta(0) < 1 - 2 S_\eta^2 (0)$. These relations are found by noting that $S_\eta(\eta_p)= - \eta_p/2$ from Eq. (\ref{eqn:SelfSimEqn}) and that $S_\eta(\eta_p) < S_\eta (0)$ where both slopes are negative. This then establishes that $S_\eta(0) > -1/\sqrt{2}$, which is proof that advancing solutions are linearly stable at all times.

\subsubsection{Transient and asymptotic stability of receding solutions $S(\eta)$}
We next examine the stability of receding solutions which are characterized by a downstream boundary value larger than the inlet condition, namely $S(\eta_B) = \textrm{const} > S(0) = 1$, and inlet slopes which are always positive, namely $S_{\eta}(0) > 0 $. In order to make use of the lemma above, we first apply a change of variable to Eq. (\ref{eqn:SLtype}) such that $G(\eta)$ is replaced by the product $S(\eta)G(\eta) >0$. Straightforward calculation yields yet another Sturm-Liouville eigenvalue equation with weighting factor $S^2(\eta)$:
\begin{equation}
\left(\frac{\mathcal{M} + \mathcal{M}^\dag}{2}\right) G(\eta,\tau) =
- \lambda S^2(\eta) G(\eta,\tau) ,
\end{equation}
where
\begin{equation}
\left (\frac{\mathcal{M} + \mathcal{M}^\dag}{2} \right)
= \frac{\partial}{\partial \eta} \left(\frac{S^3}{2} \frac{\partial}{\partial \eta}\right) - \frac{S}{4}\left(\eta S_\eta + S \right) .
\end{equation}
The lemma then applies here, with an adjustment for the weighting factor: $\lambda_{\textrm{min}} >(-Q/S^2)_{\textrm{min}}$. This then yields the result
\begin{align}
\lambda_{\textrm{min}} > & \min \left\{ \frac{1}{4 S}\left(\eta S_\eta + S \right) \right\} \,\,\,\, \textrm{or likewise} \\
- \lambda_{\textrm{min}} < &\max \left\{ - \frac{1}{4 S}\left(\eta S_\eta + S \right) \right\} .
\label{eqn:Qminreceding}
\end{align}
Since for receding solutions all terms in the expression $(\eta S_\eta + S)/S$ are strictly positive, then $\lambda_{\textrm{min}}$ is strictly positive and therefore $\omega_{\textsf{max}}$ is strictly negative. This demonstration establishes that receding solutions are transiently and asymptotically linearly stable to any infinitesimal disturbances $G$ satisfying homogeneous Dirichlet boundary conditions.

\subsection{Numerical results and comparison to analytic bounds}
\label{sec:numerics}
\begin{figure}
\centering
\includegraphics[scale=1.0]{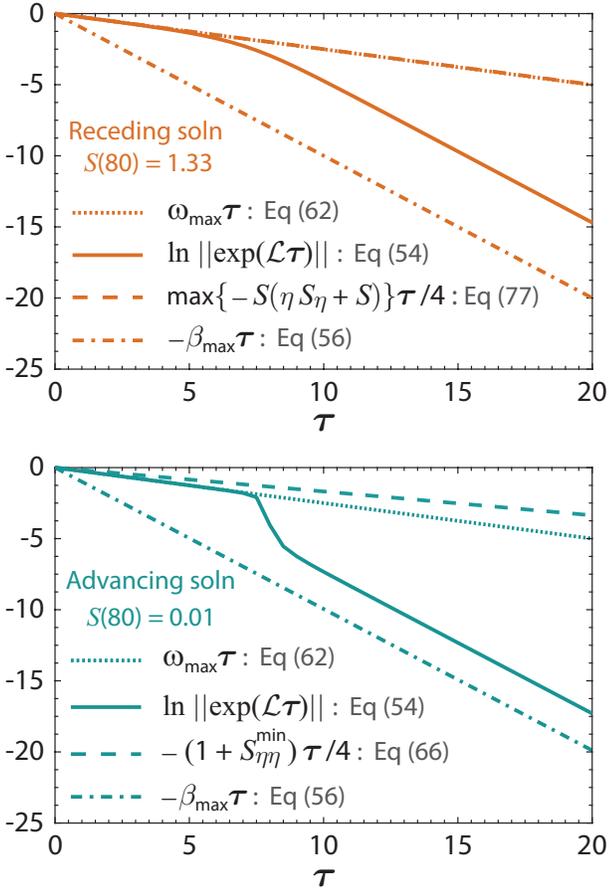}
\caption{(Color online)  Analytic bounds and numerical results confirming transient and asymptotic stability of representative receding and advancing self-similar states shown in Fig. \ref{fig:selfsimilar_solutions}. (a) Results for receding state subject to Dirichlet condition $S(\eta =80) = 1.33$. (b) Results for advancing state with Dirichlet condition $S(\eta =80) = 0.01$.}
\label{fig:numerical_stability}
\end{figure}

Our numerical results for the operator exponential governing transient growth are next compared to analytic bounds derived above. Derivatives were constructed using second-order finite differences. (In discrete form, these operators are simply square matrices.) Homogeneous Dirichlet boundary conditions were enforced by directly reducing the operator matrices. The numerical abscissa $\omega_{\textrm{max}}$ was obtained by identifying the smallest eigenvalue of the corresponding reduced operator matrix $(\mathcal{L} + \mathcal{L}^\dag)/2$ according to which $\omega_{\textrm{max}}= - \lambda_{\textrm{min}}$. The spectral abscissa $\beta_{\textrm{max}}$ was obtained by identifying the smallest eigenvalue of the reduced matrix $\mathcal{L}$. The quantity $\ln(\|\exp(\mathcal{L} \tau)\|)$, representing the maximum instantaneous disturbance amplification, was obtained from the operator norm $\|\cdot\|$ given by the matrix maximum singular value. Shown in Fig. \ref{fig:numerical_stability} are the numerical results for these three quantities plotted alongside the analytic bounds for $-\lambda_{\textrm{min}}$ given by $\max \{-(\eta S_\eta + S)/(4S)\}$ for receding solutions and $\max \{-(1 + S_{\eta \eta})/4\} = -(1 + S^{\textrm{min}}_{\eta \eta})/4$ for advancing solutions.

As predicted, the values of $\ln \|\exp(\mathcal{L}\tau)\|$ for all times $\tau$ fall  intermediate between the analytic upper bound, namely $\max \{-(\eta S_\eta + S)/(4S)\}$ for receding solutions or $-(1 + S^{\textrm{min}}_{\eta \eta})/4$ for advancing solutions, and the analytic lower bound given by $-\beta_{\textrm{max}}$. The numerical results also confirm that as $\tau \rightarrow \infty$, the slope of the function $\ln \|\exp(\mathcal{L}\tau)\|$ exactly equals the value $-\beta_{\textrm{max}}$, as must be the case since the asymptotic decay rate of disturbances is dictated by the eigenvalue of $\mathcal{L}$ with maximum real part.

\begin{figure}
\centering
\includegraphics[scale=1.0]{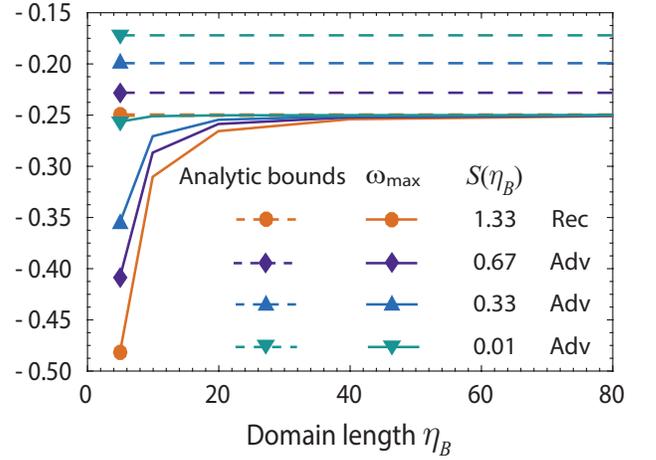}
\caption{(Color online)  Results of convergence studies as a function of the domain length $\eta_B$. The labels \textsf{Rec} and \textsf{Adv} refer respectively to receding and advancing self-similar solutions shown in Fig. \ref{fig:selfsimilar_solutions}. The quantity $\omega_\textsf{max}$ is defined in Eq. (62). The analytic bound for the receding solution is given by $\max {-(\eta S_\eta + S)/(4S)}$ defined in Eq. (77). The analytic bounds for the advancing solutions are given by $(1 + S^{\textrm{min}}_{\eta \eta})/4$ defined in Eq. (66).}
\label{fig:numerical_convergence}
\end{figure}

The numerical results in Fig. \ref{fig:numerical_stability} (and similar studies not shown of other receding and advancing states) reveal that there is no transient nor asymptotic growth of disturbances. Furthermore, the suppression of disturbances as quantified by $\ln \|\exp(\mathcal{L}\tau)\|$ asymptotes to the eigenvalue  $-\beta_{\textrm{max}}$ already at early times $\tau < 9.0$. The sudden drop off in the value $\ln \|\exp(\mathcal{L}\tau)\|$ observed near $\tau = 8$ in Fig. \ref{fig:numerical_stability}(b) simply reflects onset of increased damping of disturbances incurred in the thin precursor film. All results presented were computed using spatial domains $0 \leq \eta \leq \eta_B$ where $5 \leq \eta_B \leq 80$ with fixed mesh size ranging from 0.05 to 0.005. As shown in Fig. \ref{fig:numerical_convergence}, a domain length of 80 was sufficiently long to ensure numerical convergence irrespective of mesh size.

\section{Conclusion}
\label{sec:conclusion}
In this work, we have examined the global and linear stability of solutions describing inertia-free flow of a thin wetting Newtonian film within an open slender triangular groove of constant cross-section with constant liquid contact angle. The study is limited to flow states for which the interface function representing the local film thickness is always strictly positive thereby ruling out rupture states. The slender approximation (also known as the lubrication or long wavelength approximation) ensures that to order $\varepsilon^2$ the local value of the fluid pressure depends only on the local film thickness whose interface shape in the plane normal to the streamwise flow represents a segment of a circle. This approximation simplifies the governing equation of motion, which is described by a second order nonlinear diffusion equation.

The relevant base states for stability analysis include stationary interface states $H_S(Z)$ and self-similar states $S(\eta)$ which adopt the (dimensionless) Washburn scaling $\eta = Z/T^{1/2}$. The stationary states allow for various boundary conditions imposed at the ends of the channel including Dirichlet-Dirichlet, Dirichlet-Neumann (i.e. flux condition), Neumann-constant volume or Dirichlet-constant volume. The self-similar states are volume non-conserving and result from application of Dirichlet boundary conditions. By exploiting an analogy with other gradient flow equations and examining the asymptotic behavior of the corresponding Lyapunov function, it was shown that (strictly positive) stationary interface solutions represent exponentially stable equilibrium points. Disturbances of any type therefore decay away at least exponentially fast to restore the system back to the initial stationary state. The second important result is that advancing, uniform and receding self-similar states satisfying Washburn dynamics are both transiently and asymptotically linearly stable to infinitesimal perturbations. This finding required implementation of a generalized non-modal linear stability analysis since the inhomogeneous nature of the self-similar states naturally gives rise to non-normal disturbance operators.

These two results highlight the reasons why the interface states describing either transient flow (self-similar states) or asymptotic steady state flow (stationary states) within slender open triangular grooves tend to be so stable against perturbations so long as the liquid always wets the channel sidewalls. This result, however, requires that the  slender channel approximation be everywhere satisfied, which in turn sets an upper bound on the interface slope, namely $(\partial h/\partial z)^2 \sim (d/L)^2 \sim O(\varepsilon^2)$. Alternatively, this constraint can be viewed as a low-pass filtering requirement which prevents low-frequency modes or disturbances from ever generating  high-frequency modes. In particular, neither the base states nor disturbances applied to these base states can trigger streamwise capillary waves. Consequently, the results obtained in this study cannot address effects such as interface leveling in thin films \cite{OA63}. As noted by Yang and Homsy \cite{YH06}, flow caused by streamwise curvature can also not be neglected in the limit $(\theta + \alpha) \to \pi/2$.

The model investigated in this work also relies on the assumption that the liquid contact angle $\theta$ is a constant equal to the equilibrium contact angle, an assumption known to break down for sufficiently large flow speeds. A fully numerical model incorporating a velocity dependent contact angle \cite{BJ89,JB90} can certainly be developed, but such an extension precludes analytic solutions for base states and therefore development of analytic bounds pertinent to stability. We are currently extending this model in a new promising direction by including effects due to substrate axial curvature. We hope that the stability findings reported in this work as extended to curved substrates will ultimately provide a full analytic and numerical treatment particularly useful to the design of novel propellant management systems for various space applications.

\begin{acknowledgments}
The authors gratefully acknowledge financial support from a 2016 NASA/Jet Propulsion Laboratory President's and Director's Fund and a 2017 NASA Space Technology Research Fellowship (NCW).
\end{acknowledgments}

\appendix
\section{Boundedness of constant $\mathcal{C}$ in Eq. (\ref{eqn:constC})}
\label{sec:A1}
We first note that the constant $\mathcal{C}$ given by Eq. (\ref{eqn:constC}), namely
\begin{equation}
\mathcal{C} = \min_{T,Z} \!
\left \{\frac{2 H^3 + 4 H^2 H_S + 6 H H^2_S+ 3 H^3_S}{5(H^2 + H H_S + H^2_S)^2} \right \} .
\end{equation}
satisfies the condition $\mathcal{C} \geq 0$ for $H(Z,T)>0$ and $H_S(Z)>0$. We can eliminate the troublesome possibility $\mathcal{C} = 0$, however, as follows. Assuming $H_S(Z)$ and $H(Z,T)$ remain bounded, then the function in brackets is observed to decrease monotonically with $H$ and $H_S$ since the derivatives
\begin{align}
&\frac{\partial}{\partial H} \left(\frac{2 H^3 + 4 H^2 H_S + 6 H H^2_S+ 3 H^3_S}{5(H^2 + H H_S + H^2_S)^2} \right) = \\
&-\frac{2 H \left( H^3 + 3 H^2 H_S + 6 H H_S^2 + 5 H_S^2 \right)}{5\left( H^2 + H H_S + H_S^2 \right)^3} <0
\end{align}
and
\begin{align}
&\frac{\partial}{\partial H_S} \left(\frac{2 H^3 + 4 H^2 H_S + 6 H H^2_S+ 3 H^3_S}{5(H^2 + H H_S + H^2_S)^2} \right) = \\
& -\frac{3 H_S^2 \left( H^2 + 3 H H_S + H_S^2 \right)}{5\left( H^2 + H H_S + H_S^2 \right)^3} <0 \,.
\end{align}
The smallest possible value of the bracketed term in $\mathcal{C}$ is attained when that ratio is evaluated at the maxima of $H_S(Z)$ and $H(Z,T)$ (which don't necessarily occur at the same point $Z$). It then follows that
\begin{align}
\mathcal{C} \ge \left(\frac{2 H^3 + 4 H^2 H_S + 6 H H^2_S+ 3 H^3_S}{5(H^2 + H H_S + H^2_S)^2} \right)
\, ,
\end{align}
where $H$ and $H_S$ are evaluated at their respective maxima.
Therefore, so long as $H(Z,T) < \infty$, $\mathcal{C} > 0$.

A related issue arises in the context of the stability arguments presented in Section \ref{sec:nonlinstab}. It could be the case that although the integrated value of the free energy defined in Eq. (\ref{eqn:F(H)}) remains bounded at all times, the transient function $H(Z,T)$ might potentially diverge at finite times depending on the boundary conditions imposed. We eliminate the possibility $H(Z,T) \to \infty$ by appealing to the governing equation of motion, which essentially describes the intrinsic diffusive behavior of the $H^2$.

Consider first the behavior of Eq.(\ref{eqn:InterfaceEqn}) given by
\begin{equation}
\frac{\partial H}{\partial T} = \left( \frac{\partial H}{\partial Z} \right)^2 + \frac{H}{2} \frac{\partial^2 H}{\partial Z^2} \,
\label{eqn:Heqn}
\end{equation}
in the vicinity of local extrema. Any local maximum of $H$ will satisfy $\partial H/\partial Z = 0$, $\partial^2 H/\partial Z^2 \le 0$ and thus $\partial H/\partial T \le 0$, which leads to a diminishment in height due to the diffusive nature of the underlying equation of motion. Similarly, a local minimum of $H$ cannot further decrease in value. Therefore, any new extrema beyond those present in the initial condition can only be reached by possible extrema at the boundaries. For those solutions satisfying Dirichlet conditions at both endpoints of the domain, $H$ can never therefore attain a new maximum above the maximum value of the initial condition nor attain a new minimum below the  minimum value of the initial condition. This behavior then guarantees that so long as the initial condition satisfies the constraint $H(Z,T=0) >0$, then $\mathcal{C}$ will always remain bounded from below.

A Neumann, or equivalently, a flux boundary condition poses a potential problem since too large a flux for a given initial condition could potentially lead to a local drainage spot where the film thickness vanishes. To assess this case, we recast Eq. (\ref{eqn:H2Master}) not in terms of the local variable $H(Z,T)$ but the local flux $Q(Z,T) = - H^2 \partial H/\partial Z = - (1//3)\partial H^3/\partial Z^3$ by first multiplying Eq. (\ref{eqn:H2Master}) by $-H/2$ followed by differentiation by $Z$, which yields
\begin{equation}
\frac{\partial Q}{\partial T} = \frac{\partial}{\partial Z}\left( H \frac{\partial Q}{\partial Z} \right) = H \frac{\partial^2 Q}{\partial Z^2} + \frac{\partial H}{\partial Z} \frac{\partial Q}{\partial Z} \, .
\label{eqn:Qeqn}
\end{equation}
Adopting a similar argument as above, for any case requiring constant liquid volume, the prescribed fluxes at the endpoint must be equal, thereby imposing Dirichlet conditions on the flux $Q$. It then follows that $Q(Z,T)$ can never exceed the extrema present in the  initial condition $Q(Z,T=0)$, which therefore precludes $H$ from becoming arbitrarily large; hence $\mathcal{C}$ is again bounded below at all times.

For cases subject to one Dirichlet and one flux boundary condition, the only way in which  $H$ might approach infinity is if the flux boundary condition is made very large. However, this would give rise either to an arbitrarily large internal flux extremum, which is disallowed by the previous argument, or an arbitrarily large flux of liquid at the boundary subject to the Dirichlet condition. For the latter case, the divergence in $H$ at one endpoint would be driven by the flux condition at other endpoint, such that the system would approach infinite volume and the Lyapunov functional would therefore not decrease as required. Hence, such a situation cannot occur, and $H$ is therefore bounded above at all times and $\mathcal{C}$ bounded below at all times.

Finally, for the case of a Dirichlet boundary condition coupled with a requirement of constant volume, were $H$ to approach infinity at one boundary, the flux there would also diverge and be matched by identical divergence at the other end point in order to satisfy constant volume. But according to Eq. (\ref{eqn:Qeqn}), an interior flux minimum would have to undergo increase thus increasing the minimum slope of $H^3$, which would eventually violate the constraint of constant volume. Hence, in this case as well, $H$ cannot diverge at any point in time and $\mathcal{C}$ therefore remains bounded below at all times.

In conclusion, the demonstration above therefore establishes that all stationary states, irrespective of the boundary conditions imposed above, are exponentially stable in the Lyapunov sense.

%


\end{document}